\RequirePackage[2020-02-02]{latexrelease}
\documentclass[a4paper, twocolumn,
showpacs,longbibliography, accepted=2022-09-07]
{quantumarticle}
\pdfoutput=1

\usepackage[utf8]{inputenc} 
\usepackage{amsmath}
\usepackage{amssymb}
\usepackage{mathrsfs}
\usepackage{bm}
\usepackage{color}
\usepackage{xcolor}
\usepackage{float}
\usepackage{amsthm}
\usepackage{appendix}
\usepackage{subfig}
\usepackage{url}
\usepackage{graphicx}
\usepackage[numbers,sort&compress]{natbib}

\usepackage{lipsum}

\usepackage{braket}
\usepackage{algorithm}

\usepackage{graphicx}
\DeclareGraphicsExtensions{.pdf,.jpg,}

\usepackage{todonotes} 
\usepackage{hyperref}

 \newcommand{\red}[1]{#1}

 \newcommand{\ssout}[1]{}
 \newcommand{\esout}[1]{}

\floatname{algorithm}{Protocol}

\newcommand{\kb}[1]{\ket{#1}\bra{#1}}

\newcommand{\bk}[1]{\braket{#1|#1}}

\newcommand{\re}{\mbox{Re}}
\newcommand{\pr}{\mbox{p}}

\newtheorem{theorem}{Theorem}

\begin{document}

\author{Francesco Massa}
\affiliation{University of Vienna, Faculty of Physics, Vienna Center for Quantum Science and Technology (VCQ), Boltzmanngasse 5, Vienna A-1090, Austria}
\author{Preeti Yadav}
\affiliation{Instituto de Telecomunica\c{c}\~oes, 1049-001 Lisbon, Portugal}
\affiliation{Departamento de Matem\'{a}tica, Instituto Superior T\'{e}cnico, Universidade de Lisboa, Av. Rovisco Pais, 1049-001 Lisboa, Portugal}
\author{Amir Moqanaki}
\affiliation{University of Vienna, Faculty of Physics, Vienna Center for Quantum Science and Technology (VCQ), Boltzmanngasse 5, Vienna A-1090, Austria}
\author{Walter O. Krawec}
\affiliation{Computer Science and Engineering Department, University of Connecticut, Storrs, CT 06269, USA}
\author{Paulo Mateus}
\affiliation{Instituto de Telecomunica\c{c}\~oes, 1049-001 Lisbon, Portugal}
\affiliation{Departamento de Matem\'{a}tica, Instituto Superior T\'{e}cnico, Universidade de Lisboa, Av. Rovisco Pais, 1049-001 Lisboa, Portugal}
\author{Nikola Paunkovi\'c}
\affiliation{Instituto de Telecomunica\c{c}\~oes, 1049-001 Lisbon, Portugal}
\affiliation{Departamento de Matem\'{a}tica, Instituto Superior T\'{e}cnico, Universidade de Lisboa, Av. Rovisco Pais, 1049-001 Lisboa, Portugal}
\author{Andr\'{e} Souto}
\affiliation{Instituto de Telecomunica\c{c}\~oes, 1049-001 Lisbon, Portugal}
\affiliation{LASIGE, Departamento de Inform\'atica, Faculdade de Ci\^encias, Universidade de Lisboa, 1749-016 Lisboa, Portugal}
\author{Philip Walther}
\affiliation{University of Vienna, Faculty of Physics, Vienna Center for Quantum Science and Technology (VCQ), Boltzmanngasse 5, Vienna A-1090, Austria}

%



\title{Experimental semi-quantum key distribution with classical users}
\begin{abstract}
Quantum key distribution, which allows two distant parties to share an \red{unconditionally} secure cryptographic key, promises to play an important role in the future of communication. For this reason such technique has attracted many theoretical and experimental efforts, thus becoming one of the most prominent quantum technologies of the last decades. The security of the key relies on quantum mechanics and therefore requires the users to be capable of performing quantum operations, such as state preparation or measurements in multiple bases. A natural question is whether and to what extent these requirements can be relaxed and the quantum capabilities of the users reduced. Here we demonstrate a novel quantum key distribution scheme, where users are fully classical. In our protocol, the quantum operations are performed by an untrusted third party acting as a server, which gives the users access to a superimposed single photon, and the key exchange is achieved via interaction-free measurements on the shared state. We also provide a full security proof of the protocol by computing the secret key rate in the realistic scenario of finite-resources, as well as practical experimental conditions of imperfect photon source and detectors. Our approach deepens the understanding of the fundamental principles underlying quantum key distribution and, at the same time, opens up new interesting possibilities for quantum cryptography networks.

\end{abstract}

\maketitle
\section{Introduction}

Quantum key distribution (QKD) is a technique that allows two distant parties, traditionally called Alice and Bob, to exchange a cryptographic key in an information-theoretic secure way. This means that the security of the key relies on information theory and cannot be broken even by an eavesdropper with unlimited resources. 

The first QKD proposal was the BB84 protocol, introduced by Bennett and Brassard in 1984~\cite{QKD-BB84} (subsequently, Ekert introduced the E91 protocol in 1991~\cite{QKD-E91}), which was proven secure several years later~\cite{QKD-BB84-rate1,QKD-renner-keyrate,QKD-Winter-Keyrate}. Since then, much progress, both theoretical and experimental, has been made in the field. The practicality of this technology is underlined by numerous experimental and even commercial endeavors, supporting its development~\cite{QKD-survey,she:17,raz:19,xu:19}. 

Most QKD protocols require Alice or Bob to share a quantum state, or a direct quantum channel, and to perform quantum operations, i.e., operations on quantum bits (qubits) that do not have any counterpart in classical communication, such as generation or measurement in multiple bases. On the other hand, it is known that if both parties are restricted to classical communication, unconditional security is unachievable for the key distribution problem. It is therefore relevant for a fundamental understanding of QKD to investigate how quantum the users' operations and resources need to be in order to achieve information-theoretic security.

A first step in this direction was made by introducing the \emph{semi-quantum} model of cryptography in 2007 by Boyer et al.~\cite{SQKD-first}. In this model, at least one party must be ``classical'' in nature, i.e., restricted to a limited set of operations on qubits, namely measuring and/or preparing qubits in a single basis (usually the computational ($Z$) basis $\{\ket{0}, \ket{1}\}$), or simply disconnecting from the quantum channel by allowing any received quantum state to reflect back to the sender. The use of ``classical'' in this terminology is due to the fact that orthogonal quantum states from a single measurement basis and states of classical systems are both fully distinguishable. The other parties may be classical or quantum (naturally, at least one party must be quantum) with a ``quantum'' user having the ability to perform any quantum operation on qubits allowed by the laws of physics. In the subsequent proposal~\cite{SQKD-second}, permuting or reordering the incoming qubits using delay lines was considered as another classical operation. Nevertheless, although one can indeed argue that permuting physical systems is inherently classical operation, doing so, especially in photonic applications, is with the current technology far more infeasible than any quantum operation used in cryptographic protocols. Also, preparing and detecting qubit states, albeit in a single basis, is technologically non-trivial. 

Further development has shown that Alice's operations can be as limited as Bob's, provided that a third party distributes entangled photons to the users and performs measurements in different bases~\cite{SQKD-MultiUser,liu2018mediated}. Such a scheme, referred to as a mediated SQKD protocol, allows two classical users to establish a shared secret key with one-another, using the help of a quantum server which must prepare, and later measure, quantum bits. However, this quantum server need not be trusted, and in fact could be an all-powerful adversary. 
Security was proven, but again, only for the perfect-qubit scenario~\cite{SQKD-MultiUser}.

Since that original mediated-SQKD protocol, there have been several advances both in new protocol design and in new security proof methods.  A main research goal in this field is to develop new protocols which further reduce the requirements placed on either the end-users or the server (or both).  In terms of reducing the complexity of the end-users, a protocol which did not require users to measure was proposed in~\cite{liu2018mediated} (however, attacks against the protocol were later discovered in \cite{zou2020three}).  On the other hand, in \cite{lin2019mediated,chen2021efficient}, protocols were developed which reduced the server's requirements.  Namely, in \cite{lin2019mediated} the server need only send single qubit states to users but later requiring a Bell measurement.  In \cite{chen2021efficient} single qubits were used, both in the initial preparation stage and in the subsequent server measurement, however a cycle topology was required.

Beyond reducing end-user or server requirements, another avenue of research in this area is in improving either efficiency or noise tolerance of the protocol (or both) and in developing new security proof methods.  In \cite{krawec2019multi} a new multi-mediated model was introduced which could improve noise tolerance at the cost of efficiency, while in \cite{guskind2022mediated} a new protocol was introduced which improved efficiency (though at the cost of noise tolerance).  

Most SQKD protocols up to this point have been theoretical in nature, and assume perfect qubit channels, i.e., no photon loss or multi-photons are permitted for their security to be valid. A SQKD protocol immune to such imperfections was described recently in~\cite{boyer2017experimentally} and was proven to be robust, meaning that any attack which causes an adversary to gain non-zero information on the key, necessarily creates a disturbance that may be detected with non-zero probability. A second such protocol was proposed in~\cite{krawec2018practical}, though there security was only proven against a few specific attacks. However, no full proof of security yet exists for these protocols and so, their key rates and noise tolerances are still unknown. 

In general, while numerous SQKD protocols have been proposed in the last decade~\cite{iqbal2019semiquantum}, information-theoretic proofs of security were developed only for a few of them ~\cite{SQKD-MultiUser, krawec2019multi, guskind2022mediated, krawec2015security, zhang2018security} and always in the ideal scenario of perfect qubits, ideal devices and infinite resources in the asymptotic regime.

In this work, we propose a novel SQKD protocol in the mediated model, allowing two classical users to share a secret key using the help of an untrusted, potentially adversarial, quantum server. In particular, our protocol requires Alice and Bob to perform two classical operations only, the detection or reflection of a single photon, and hence places even fewer restrictions on the users than prior protocols of this nature, by requiring only a single photon measurement and no state preparation. We are the first to show that such minimal requirements, on the part of the users, is sufficient to generate a secret key. The server's complexity is also reduced compared to prior work, needing only to prepare and measure single qubits. Furthermore, as first for mediated SQKD research, we conduct an information theoretic proof of security of the protocol assuming practical devices, whereas prior work in mediated SQKD was restricted to perfect qubit scenarios, and compute the secret key rate in the finite key setting. Finally, we experimentally demonstrate our protocol under real-life conditions and evaluate the secret key rate by using the results from actual devices. Our methods here may also be broadly applicable to other multi-user (S)QKD protocols in practical settings.

\section{The Protocol}\label{sec:protocol}
Our protocol involves three parties: two classical users, Alice and Bob, whose aim is to exchange a secret cryptographic key, and an untrusted, potentially adversarial, quantum server, which provides the quantum resources for this purpose. Furthermore, we assume that Alice and Bob can communicate through a classical authenticated channel and that the server can send unauthenticated classical messages to the users. In the description below, we discuss the protocol for single photons for simplicity, and also they are the most practical for QKD applications (though our security analysis will also take into account realistic multi-photon sources).

\begin{figure}[t]
  \centering
  \includegraphics[width=\linewidth]{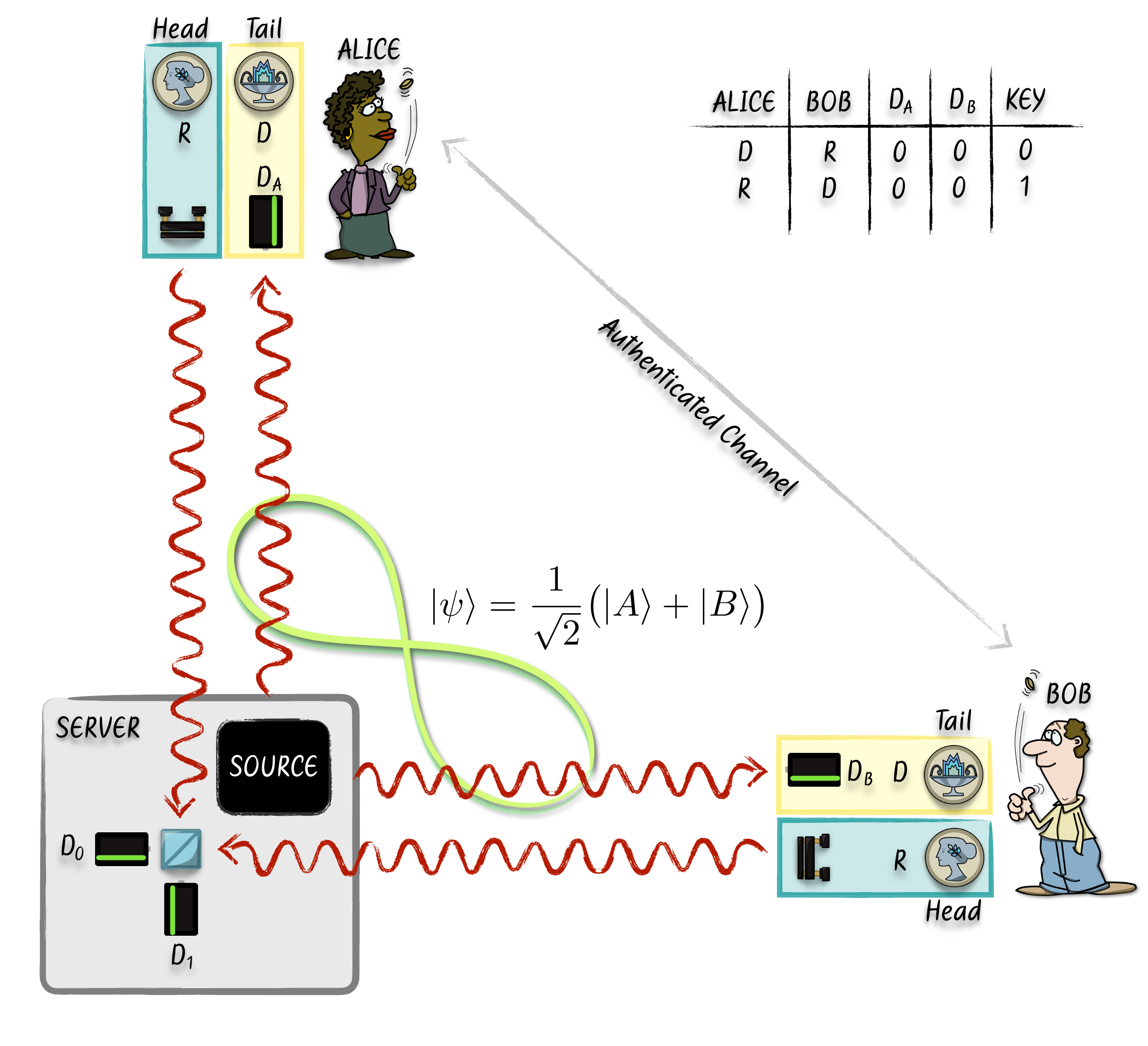}
\caption{\footnotesize \textbf{The QKD protocol with classical users.} A quantum server sends single photons in superposition to the users at predetermined regular intervals, which constitute the rounds of the protocol. For each round, Alice
and Bob randomly choose between "detect" (D) and "reflect" (R). The photons reflected back to the server impinge onto a beam splitter at whose outputs two detectors D0 and D1 are placed. When both Alice and Bob reflect the received photon, single-photon interference occurs at the beam splitter and only detector D0 clicks. If only one of the users chooses to detect the photon without registering any detection event, interference is suppressed and the photon has ideally 50\% probability to reach detector D1. If the server announces a detection at D1 and none of the users detected photons, a raw key bit is generated according to the table in the figure. The users can communicate through a classical authenticated channel to verify the honesty of the server and to share the necessary information for the evaluation of the secure key rate.}\label{fig:theory_fig}
\end{figure}

A sketch of the scheme is depicted in Figure~\ref{fig:theory_fig}. The server sends to Alice and Bob a single photon in a balanced superposition of their respective locations. Each user can independently choose to perform two actions: ``detect'' (\textit{D}) or ``reflect'' (\textit{R}). In the former case, the photon travels to a detector controlled by the user; in the latter, the photon is sent back to a balanced beam splitter controlled by the server, at whose output ports two detectors, D$_0$ and D$_1$, are placed. When both users choose to reflect, single-photon interference occurs at the beam splitter, with the relative phase of the two interfering photon amplitudes tuned such that only detector D$_0$ clicks. In the ideal case of perfect detection efficiency, when only one of the users chooses to detect the photon and does not find any, the photon collapses into the other user's location. This corresponds to performing an {\em interaction-free} measurement~\cite{dic:81,elz:vai:93,vai:95}, which suppresses single-photon interference at the server and allows \ssout{both detectors} \red{either detector} D$_0$ and D$_1$ to click with non-zero probability. A click at detector D$_1$, therefore, enables each user to deduce the action of the other one, thus allowing for the establishment of a raw key digit. In particular, a key digit of 0(1) is set when Alice chooses \textit{D}(\textit{R}) and Bob \textit{R}(\textit{D}). Other combinations are not considered, as they cannot result in a detection at D$_1$. Since the raw key bits are generated when the server announces a click at detector D$_1$, and {\em neither} Alice {\em nor} Bob detect a photon, no use of the authenticated channels is needed during those rounds, unlike the standard QKD protocols ~\cite{QKD-BB84,QKD-E91}. In our protocol, classically authenticated information exchange is performed only for the verification and parameter estimation rounds, which are not used to generate the raw key.

The detailed steps of the protocol are described below:

\red{  \textbf{Quantum Communication Stage: } Users repeat the following process until a sufficiently large raw key has been established (refer also to Figure \ref{fig:theory_fig}):
  \begin{enumerate}
  \item The server sends a single photon to both parties in a superposition.  Ideally this should be performed by the server sending a single photon through a beam splitter.
  \item Alice and Bob choose, independently and randomly, between two available actions: $D$ or $R$.  Since Alice and Bob are completely classical, the detection results only give them information as to whether or not there is a photon at their respective detector $D_A$ or $D_B$.  Their actions determine their raw key bit for this round, namely:
    \begin{itemize}
    \item \textbf{Alice: } If Alice chose $D$, she will record a raw key-bit of $0$; otherwise, if she chose $R$, she will record a raw key-bit of $1$.
      \item \textbf{Bob: } Bob's encoding is opposite that of Alice; namely if he chose $D$ he will record a raw key bit of $1$ and, otherwise, a raw key bit of $0$ if he chose $R$.
    \end{itemize}
  \item The server measures the photon coming from Alice/Bob and announces the following results: ``$0$'' if the server's detector $D_0$ clicks; ``$1$'' if detector $D_1$ clicks; ``$v$'' if no detector clicks; or ``$m$'' if more than one detector clicks.  Ideally, this measurement should be performed by the server completing a (folded) Mach-Zehnder interferometer as shown in Figure \ref{fig:theory_fig}.  Note that the last case can arise due to experimental imperfections or the action of an adversary.
  \item Alice and Bob perform a minimal sifting step whereby they will keep the round only if the following two conditions are met:
    \begin{itemize}
    \item The server announces the message ``$1$''
      \item \emph{and} Alice and Bob both did \emph{not} detect a photon if they chose to measure.
    \end{itemize}
    All other events will cause the round to be discarded.  Note that, for this, Alice and Bob must announce whether they detect a photon or not.  In the event parties choose $R$ they will, by default, announce that they did not detect a photon.
  \end{enumerate}

  \textbf{Sampling Stage: } Users will communicate, through an authenticated channel, their actions and measurement outcomes (if applicable) for a randomly chosen subset of the rounds performed above.  This is done to verify the honesty of the server and/or the presence of an adversary.  More specifically, these statistics, as discussed below, will be used to determine a bound on the overall key-rate of the protocol.

  \textbf{Post Processing Stage: } After performing the above sampling process and discarding those rounds chosen for sampling, users will perform a standard error correction protocol and privacy amplification protocol resulting in the final secret key of the system.  For information on these standard processes, the reader is referred to \cite{QKD-survey}.}

\red{It is not difficult to see that, if the server is honest, the protocol is correct.  Namely, the only time the server should ever send the message ``$1$'' is when Alice and Bob choose opposite actions (thus resulting in a correlated raw key bit since their encoding operations are opposites of one another).  We show later that the protocol can lead to a secure secret key even if the server is adversarial.}


\section{Key generation and parameter estimation}\label{sec:verification}

In this section, we discuss the events when raw key bits are generated and the parameter estimation procedure (for details see Appendices \ref{sec:key_extraction}, \ref{sec:ideal}, and \ref{sec:bound_entropy}).

Let $N$ be the total number of successful rounds in the protocol, i.e., whenever the server announces a message from ``$v$'', ``0'', ``1'' or ``$m$''. At the end of $N$ rounds, Alice and Bob communicate with each other over an authenticated classical channel to proceed to, first, the verification procedure, and then, to estimate the parameters to eventually share a secret key among them. Note that the server is bound to announce the same results to both Alice and Bob, since it can easily be checked by the users when they communicate over an authenticated channel. Therefore, upon having all the indexed results from the server, each user compares it with their own action. During the rounds when the server announced ``1'', when a user either reflected, or detected vacuum, only then we say that the user's action is ``consistent'' with the server's result, and no information is sent to the other client. Otherwise, any of the users detecting inconsistency announces it to the other one and the corresponding round is discarded from the rounds for key-generation. Such inconsistencies could be due to receiving a click in their detectors, or receiving clicks even when they reflected due to the failure in the switch used by them to change between the actions reflect ($R$) or detect ($D$).

Therefore, when the server announces ``1'' and both users' actions are consistent with such outcome, then a raw key digit is generated. This occurs on total of $N_{raw} = \pr (1)N$ rounds, where $\pr (1)$ denotes the probability that the server announces ``1'' and none of the users detect any click(s). The cases when the server announces ``1'' and both users reflected or both detected vacuum determine errors in the key.

Note that in the majority of QKD protocols (for instance, BB84), even the very first set of keys shared by Alice and Bob requires them to communicate over an authenticated channel. On the contrary, the first set of shared key in our protocol does not require any communication between the users, but only the message ``1'' from the server.

Alice and Bob choose each action ($R$ or $D$) independently at random, with probability $1/2$. Thus, the cases when the key can potentially be generated occur with probability $1/2$. In those cases, in ideal conditions, there is a probability of $1/2$ that the photon collapses into the location of the user that reflects. Finally, the reflected photon has at best a further probability of $1/2$ to come out from the beam splitter at the output of detector D$_1$. Therefore, $\pr (1)$ is at best $1/8$, which is further reduced by experimental imperfections, eavesdropping or the action of an adversarial server.

For the rest of $(1-\pr(1)) N$ rounds, the users exchange the information of their actions and detection results over the classical channel in order to estimate the parameters necessary for the establishment of a secret key between them. Note that it is enough that only one user, say Alice, performs the verification with the information received from the other. This allows for a reduction of the communication complexity. In addition to his action choices and results for the $(1-\pr(1)) N$ rounds, Bob can also send the messages announced by the server over all the rounds. Alice will proceed with parameter estimation only if {\em all} of Bob's messages match with hers. 

Using the information received from Bob for the $(1-\pr(1)) N$ rounds, Alice can perform an indirect estimation method to evaluate the probability of exchanging a key bit, $\pr_{key}$, and the probability of error on the key, $\pr_{err}$, without the need to discard any key bit. A drawback of this indirect estimation is that $\pr_{key}$ and $\pr_{err}$ are obtained from other directly-measured quantities, therefore, due to error propagation, their uncertainty is higher. Alternatively, the users can exchange full information about their actions for a randomly chosen fraction $\tau$ of $N_{raw}$ rounds to directly estimate the necessary probabilities. However, in the direct estimation, the uncertainty of the final probabilities depends on the size $\tau$ of the considered sub-sample. The choice of which method to use, therefore, depends on the experimental parameters and the length of the raw key.

\section{Experimental implementation}\label{sec:experiment}

\begin{figure*}[t]
\centering
\includegraphics[scale=0.4]{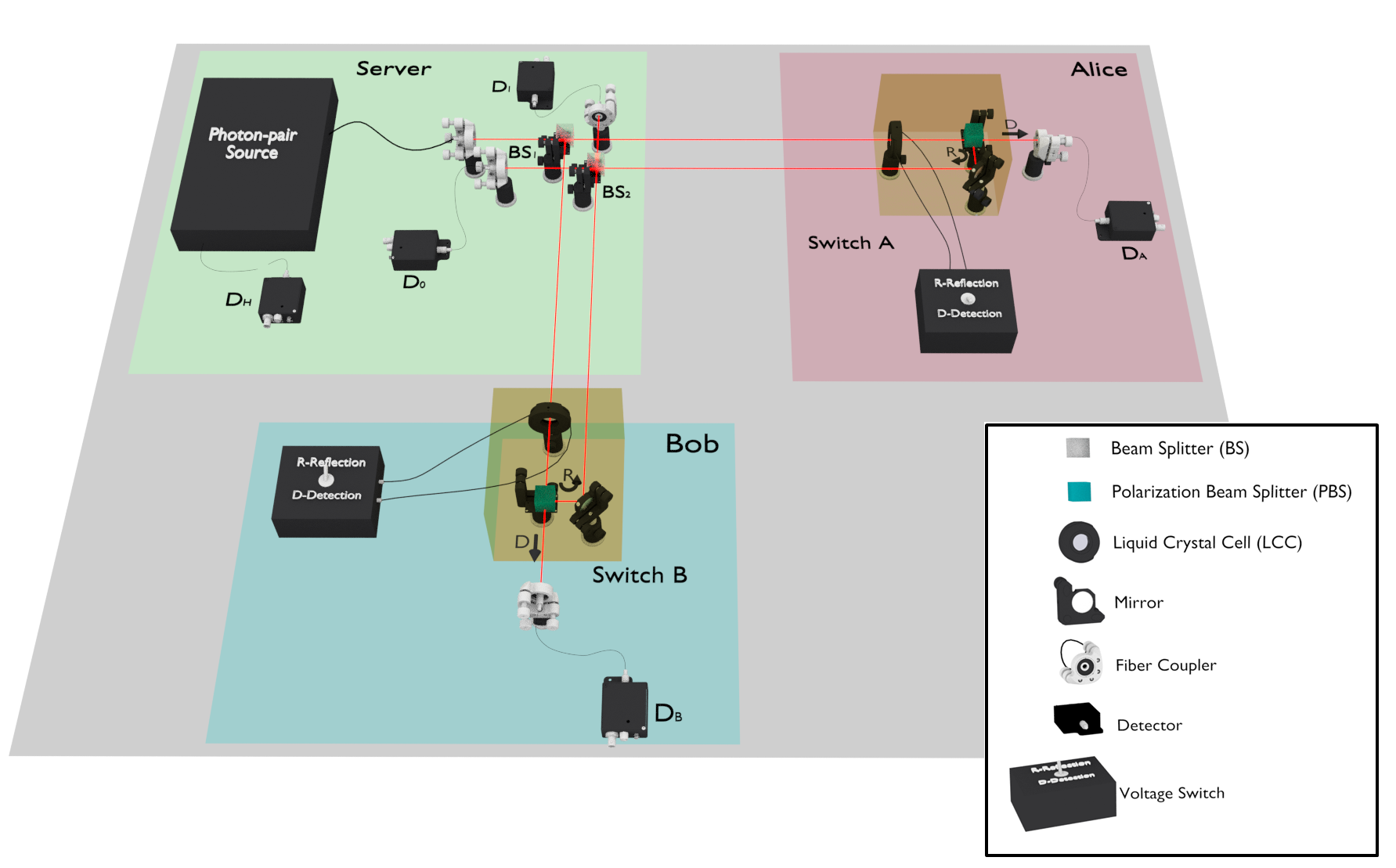}
\caption{\footnotesize \textbf{Experimental set-up.} The regions of space occupied by Alice, Bob and the server are respectively marked in red, blue and green, whereas the path of the photons is indicated by red lines. The server uses a heralded single-photon source and a beam splitter (BS$_1$) to produce the superposition state that is sent to Alice and Bob. Each of the users controls a switch, composed of a liquid-crystal cell (LCC) at 45$^{\circ}$, a polarization beam splitter (PBS) and a mirror. By switching the voltage of the LCC, the users can choose to steer the photon to a detector (D) or reflect it back to the server (R). The server collects the reflected photons at a second beam splitter (BS$_2$), where single-photon interference takes place in case both users choose to reflect. The server records the detections at D$_0$ and D$_1$ and announces the results to the users via a classical channel.}\label{fig:set-up}
\end{figure*}

The experimental set-up for the implementation of the protocol is depicted in Figure~\ref{fig:set-up}. After setting its polarization to ``horizontal"(H), that is parallel to the optical table, a single photon is sent to a beam splitter that creates the superposition between Alice's and Bob's locations. Each of the users controls a liquid-crystal cell (LCC) at 45$^{\circ}$ and a polarization beam splitter (PBS). The phase retardation between the two axes of the LCC can be switched between 0 and $\pi$ by means of a voltage signal. Consequently, the photon polarization is rotated by 0$^{\circ}$ or 90$^{\circ}$, respectively. In the first case, the photon is transmitted by the PBS and steered to a fiber-coupled avalanche photo-diode (APD) for detection, D$_{\text{A}}$ or D$_{\text{B}}$; in the second case, the photon travels back to the server. The detection efficiency of D$_{\text{A}}$ and D$_{\text{B}}$ is evaluated by comparison with a fully-characterized transition-edge superconducting nanowire detector. The photons going back to the server impinge onto a second beam splitter, at whose outputs two fiber-coupled APDs, D$_0$ and D$_1$, are placed. The set-up, therefore, implements a folded Mach-Zehnder interferometer.
The phase between the two arms of the interferometer is set such that, when Alice and Bob both decide to reflect back the photon, detector D$_0$ clicks. The interferometer is passively stabilized, so that the phase is constant for about 100 s. After this time, the phase is actively re-set to the initial value by using a piezo transducer.

The single photons are provided by a source based on spontaneous parametric down-conversion (SPDC) in a 20 mm-long periodically-poled potassium tytanyl phosphate (PPKTP) crystal, which probabilistically converts a photon at $395$ nm from a continuous-wave laser into two photons at $790$ nm and with orthogonal polarizations. One photon from each produced pair is used to herald the presence of the other one, which is sent to the users. Therefore, all detections in the experiment are in coincidence with the heralding detector, D$_{\text{H}}$.
The server sets intervals of $0.5$ s, constituting the rounds of the protocol, in which Alice and Bob can decide to either detect or reflect the photons. Note that this interval can be made shorter, in the order of $10^{-8}$ s, by using ultra-fast switches and optimized bright single-photon sources \cite{lenzini2017}. At the end of each round, the server announces the result of the measurement at its detectors.
The probabilistic nature of our source implies that, in each round, multiple non-simultaneous single-photon emissions can occur. In some rounds, therefore, the total number of detections is higher than one. The output rate of the source is decreased, so that the total average number of photons sent to the users is about $0.35$ per round, in order to reduce the probability of multi-photon emissions. 

The possibility of simultaneous
multi-photon emission from the source is ruled out by the measurement of the heralded second-order correlation function at zero delay, $g^{(2)}(0)$ \cite{mandelwolf}, which should be exactly 0
for an ideal perfect single-photon source. We obtain
$g^{(2)}(0) = 0.004 \pm 0.010$, measured at a total detection
rate of about $15 \times 10^3$ photons per round (in our case $0.5$ s) and a pump power of $7$ mW. Our value of $g^{(2)}(0)$ is comparable
with the lowest ones obtained in quantum optics
experiments \cite{eisaman2011}.

\section{Security analysis}\label{sec:security}

%
%

We prove security of our protocol under the following assumptions:
\begin{itemize}
    \item[1.] The server may be compromised by the adversary.  In particular, it may prepare an arbitrary initial state and perform an arbitrary quantum operation on the returning signals (both subject to the other constraints listed below).  Due to this assumption, we must only analyze the case of a single adversary, namely the server, and any third party adversary’s attack may be absorbed into this adversarial server’s attack strategy (to the advantage of the adversary).

\item[2.] The adversary performs collective attacks only.  That is, the adversary attacks by using an identical attack operation at each iteration (both for the initial state preparation strategy and the final quantum operation strategy, including the message sending).  The server’s initial state may be entangled with a private quantum ancilla and the final operation may also result in a private quantum memory system.  The adversary is free to postpone measuring its ancilla until any future point in time and may even perform an arbitrary global measurement of its entire ancilla at that future point in time.

\item[3.] The attack performed by the adversary on each iteration of the protocol is not interactive/adaptive.  In particular, the adversary must prepare an initial state once at each iteration and send it to Alice and Bob. Although this initial state may consist of multiple photons, the server cannot feed a photon into Alice or Bob’s lab, and then, based on the output, immediately feed additional photons into Alice or Bob’s lab. While this seems a strong assumption, there are mechanisms to enforce its compliance as we discuss in Appendix~\ref{subsec:bomb}. Although a full analysis of interactive attacks would be very interesting, we consider it out of scope of this paper as we are primarily focused on the development, finite key analysis, and experimental demonstration of a novel mediated SQKD protocol with minimal end-user resource requirements.  We do, however, consider an interactive attack based on a ``quantum bomb'' attack in Appendix~\ref{subsec:bomb}.  

\item[4.] The initial state sent by the server consists of zero, one, or two photons prepared in an arbitrary manner.  This was done as our experimental implementation consisted of a negligible probability of producing three or more photons.  It is also an enforceable condition if Alice and Bob used cascading interferometers to ensure the state, with high probability, does not contain more than two photons.  Our proof methodology, however, can be extended to consider the three or more photon case (assuming the attack is non-adaptive in this round as discussed above) if required. While we do not work out the exact algebra in this paper for that case, we do consider a particular multi-photon attack with three or more photons in Appendix~\ref{subsec:noon}.

\end{itemize}

After Alice and Bob receive quantum states of some form from the server and perform their respective actions, they will receive a classical message from the sever indicating a possible measurement outcome. However, the server is under no obligation in our proof of security to report the measurement outcome honestly, or to even perform any measurement at all. On the rounds where the server announces ``1'', Alice and Bob generate the raw key of length $N_{raw}$ whenever one of them chose to detect the photon without registering any click at the detector, while the other reflected. Note that due to experimental imperfections and eavesdropping (or server's dishonesty), server can announce ``1'' even if both agents reflected, or both detected vacuum, in which case they do not share the same raw key and the error is introduced. As mentioned before, from the raw key of size $N_{raw}=\pr(1) N$, Alice and Bob may choose to use a (small) subset of size $\mu=\tau N_{raw}$ to directly estimate the statistics used to compute the secret key rate. The portion of the raw key remaining after parameter estimation step is called the sifted key, of the length $N_{sift} = N_{raw} - \mu$. Let the random variables $\mathcal R_A$ and $\mathcal R_B$ denote Alice's and Bob's respective sifted keys.
\red{After the quantum communication and sampling stages, it is not necessarily true that $\mathcal R_A$ and $\mathcal R_B$ are uniformly distributed or fully correlated.  It is also not necessarily true that they are completely secret.  Thus, the protocol must perform a classical post processing stage which further processes these raw key strings through error correction (to ensure they are perfectly correlated with high probability) and privacy amplification (which ensures that Eve's ancilla is independent of the final secret key.}

\red{The security level of the key shared between Alice and Bob is given by parameter $\epsilon$, which quantifies how uncorrelated the secret key is from Eve or a dishonest server.  More formally, from \cite{renner2008security,QKD-renner-practical}, one should have:}
\begin{equation}\label{eq:PA}
  \left|\left|\rho_{KE} - \frac{I_K}{2^\ell}\otimes\rho_E\right|\right| \le \epsilon,
\end{equation}
\red{where $\rho_{KE}$ is the classical-quantum state modeling the secret key (after error-correction and privacy amplification) and Eve's ancilla, while $I_K/2^\ell\otimes\rho_E$ is an ideal uniform random key of size $\ell$-bits independent of Eve.  }
The security criterion requires $\epsilon$ to tend to zero as the number of rounds $N$ tends to infinity, thus obtaining perfectly secret key in the asymptotic scenario. One can compute the sifted key rate as $r'= \lim_{N\rightarrow \infty} \ell/N_{sif} = S(A|C)-H(A|B)$ using results in \cite{QKD-Winter-Keyrate}. Conditional Shannon entropy $H(A|B)$ can be easily computed using the probabilities $p_{i,j}$ of Alice and Bob establishing the raw key bit values $i$ and $j$, respectively. Further, the secret key rate is defined as $r = \ell/N = r' (N_{sift}/N)$, which is the same as the sifted key rate in the asymptotic regime: since in order to obtain good enough statistics during the verification procedure, the number $\mu$, albeit big, is still finite, we have $N_{sif} = N - \mu \approx N$, for $N \rightarrow \infty$. In the realistic case of limited resources, however, where Alice and Bob can exchange only a finite number of keys, we must take into account the imperfect parameters. Using the security criterion given by~\cite{QKD-renner-practical}, let us denote $\epsilon_{PE}$ as a given error tolerance for the parameter estimation. One can further compute $\delta$, as a function of $\epsilon_{PE}$, a confidence interval so that the observed parameters are $\delta$ close to the actual values, except with probability $\epsilon_{PE}$. Let $\epsilon$ be the desired security of the final secret key, and let $\epsilon_{EC}$ be the maximal probability that Bob computes error correction incorrectly. All of these are given by the user. Therefore, after $\mu$ rounds are used for the direct method of parameter estimation, the proportion of qubits used for estimating the secret key rate is $(\pr(1)N-\mu)/N$. Using the results shown in~\cite{QKD-renner-practical}, under the assumption of collective attacks, we have the following Theorem:

\red{\begin{theorem}\label{thm:key-rate}
  (Modified from \cite{QKD-renner-practical}): Let $\rho_{AC}^{\otimes N}$ be the state of the quantum system produced by executing the protocol $N$ times.  Then, the key-rate $r$ is bounded by:
\begin{equation}\label{eq:secure_key-rate}
r \ge \frac{\pr (1)N-\mu}{N}\left(S(A|C)_\rho - \frac{\texttt{leak}_{EC} + \Delta}{\pr (1)N-\mu}\right),
\end{equation}
where
\begin{eqnarray}
&&\Delta = 2\log_2\left(\frac{1}{2(\epsilon - \epsilon_{EC} - \epsilon')}\right) \nonumber\\
&&\quad\quad\quad\quad + \; 7\sqrt{ (\pr (1)N-\mu) \log_2(2/(\epsilon' - \epsilon_{PE}))}.
\end{eqnarray}
\end{theorem}

Above, $S(A|C)_\rho$ is the conditional von Neumann entropy of Alice's raw key bit register conditioned on the server's quantum memory system.   The value $\texttt{leak}_{EC}$ quantifies the error-correction leakage (namely, the number of classical bits exchanged between Alice and Bob during the error correction protocol).  Finally, $\epsilon$ is the desired distance to an ideal key (as in Equation \ref{eq:PA}); $\epsilon_{PE}$ is the user specified error tolerance for the parameter estimation; $\epsilon_{EC}$ is the failure probability of the error correction protocol; and $\epsilon'$ is arbitrary (chosen by the user to maximize the expression) but bounded by $\epsilon-\epsilon_{EC} > \epsilon' > \epsilon_{PE} \ge 0$.}

\ssout{and $\epsilon'$ is arbitrary (chosen by the user to maximize the expression but bounded by $\epsilon-\epsilon_{EC} > \epsilon' > \epsilon_{PE} \ge 0$). In the above expression, $S(A|C)$ is the von Neumann entropy of Alice's raw key bit register conditioned on the server's quantum memory system.}

\red{Of course, users don't have an exact description of $\rho$ needed to directly compute $S(A|C)$ above.  Thus, to actually compute the key-rate $r$, }$S(A|C)$ is minimized over all observable statistics within the given confidence interval (so that the actual statistics of the real density operator are within $\delta(\epsilon_{PE})$ of the observed statistics, except with probability $\epsilon_{PE}$).  Later, in our security proof, we will use a theorem from \cite{QKD-Tom-Krawec-Arbitrary}, stated below as Theorem \ref{thm:cq-entropy}, to actually bound the entropy $S(A|C)$.  The value $\texttt{leak}_{EC}$ represents the number of (classical) bits exchanged between Alice and Bob during the error correction. Again, using~\cite{QKD-renner-practical}, we take $\texttt{leak}_{EC}/(\pr (1) N-\mu) = (1.2) h(Q)$, where $Q=\pr_{err}/\pr(1)$ and $\pr_{err}$ is the probability to generate opposite key bits during the entire protocol. Note that $\mu$ will also be a function of $\epsilon_{PE}$, since the smaller that is, the larger $\mu$ will be.

\red{\begin{theorem}\label{thm:cq-entropy}
  (From \cite{QKD-Tom-Krawec-Arbitrary}): Let $\rho_{AC}$ be a quantum state of the form:
  \begin{equation}
    \frac{1}{N}\sum_{a=0}^2\kb{a}_A\otimes\left(\sum_{i=1}^{N}\kb{F_i^a}_C\right).
  \end{equation}
  Then, the von Neumann entropy $S(A|C)_\rho$ may be bounded by
  \begin{align*}
    S(A|C)_\rho &\ge \frac{1}{N}\sum_{i=1,N}\left(\bk{F_i^0} + \bk{F_i^1}\right)\\
    &\times\left[h\left(\frac{\bk{F_i^0}}{\bk{F_i^0} + \bk{F_i^1}}\right) - h(\lambda_i)\right],
  \end{align*}
  where
  \begin{align*}
    \lambda_i = \frac{1}{2}\left(1 + \frac{\sqrt{(\bk{F_i^0} - \bk{F_i^1})^2 + 4\textnormal {Re}^2\braket{F_i^0|F_i^1}}}{\bk{F_i^0} + \bk{F_i^1}}\right).
  \end{align*}
\end{theorem}}

\ssout{In Appendices \ref{sec:ideal} and \ref{sec:bound_entropy}, we describe in detail the procedure to find the lower bound of $S(A|C)$, the quantum-capable adversary's uncertainty on the key, for our protocol.}
\red{At a high level, our security proof involves bounding the conditional von Neumann entropy $S(A|C)$ of the system assuming an adversarial server.  This is achieved by first writing out an explicit description of the overall state's density operator (including the photons in the interferometer, the agents, and the Server/adversary).  We then show how certain important qualities of the state, namely the overlap of various ancilla vectors of the adversary, may be determined through observable statistics (such as, for instance, $\mbox{p}_{err}$).  Finally, we use Theorem \ref{thm:cq-entropy} to bound the conditional entropy and Theorem \ref{thm:key-rate} to determine a final bound on the secret key rate.  These steps are algebraically involved and so are derived in detail in the appendices.  Namely, in Appendix~\ref{sec:ideal} we derive the key rate for the ideal-qubit case.  This first stage also helps to develop the intuition of the proof used for the more complicated scenario involving practical device imperfections, presented in Appendix \ref{sec:bound_entropy}.}
Bounding $S(A|C)$ is the critical, and challenging, element of any QKD security proof. The techniques to bound this quantity developed in this work may be useful in other protocols as well.

Our security analysis takes into account the finite detection efficiencies of commercial single-photon detectors and multi-photon components in the quantum state received by Alice and Bob (see Appendix \ref{sec:key_extraction}), but does not consider other imperfections which can be used by an eavesdropper to gain information about the key. This is in general an issue for all cryptographic protocols, both classical and quantum, as it is in practice very challenging to consider all potential side channels in the security analysis \cite{sidechannelsQKD, sidechannelsQKD-1, sidechannelsQKD-2, sidechannelsQKD-3, sidechannelsclassical}. However, specific attacks can be countered by technical adaptations of the experimental set-up. As an example, let us consider the frequency dependence of the APD's detection efficiency. By sending photons at frequencies outside the detection bandwidth of the users' detectors, an eavesdropper can in fact gain information about the agents' actions while remaining completely undetected. This specific issue can be solved by employing bandpass filters that block any incoming light at undetectable frequencies. Similar strategies can be used for other degrees of freedom which the eavesdropper could exploit to prepare undetectable photons (e.g. time, spatial mode, etc.). Current photonic technology provides effective filtering systems for all these degrees of freedom \cite{filtering1, filtering2, filtering3, filtering4}, which allows the users to counter the described category of attacks at the price of a more complicated set-up and a reduction in the secret key rate.

\section{Experimental Results}\label{sec:results}
To obtain the numerical values from the lower-bounds on $S(A|C)$ and other terms from the expression~\eqref{eq:secure_key-rate} for the secret key rate, $r$, we measure the probability of the raw key generation, $\pr_{key}$, and the probability of error in the raw key, $\pr_{err}$, after $10^5$ rounds of the protocol. \red{Formally, $\pr_{key}$ is defined to be the probability of Alice and Bob not rejecting a round, while $\pr_{err}$ is the probability that, conditioned on a raw key bit being distilled, that the raw key bit contains an error (e.g., Alice has a $0$ while Bob has a $1$).  Note that $10^5$ rounds is not sufficient to actually produce a secret key through this protocol under these operating conditions as our later evaluations show; however, it is sufficient as a proof of concept to gather experimental statistics and evaluate what the key-rate \emph{would be} had we continued the experiment for a longer duration.}

The values of $\pr_{key}$ and $\pr_{err}$ are evaluated in three different ways: direct estimation over the full data set, direct estimation over a randomly chosen subset of $10^4$ rounds and indirect estimation. In the direct estimation, the users sacrifice a part of the raw key for verification procedure (see Appendix \ref{subsubsec:direct_estimation} for details). In the indirect estimation, discussed in detail in Appendix~\ref{subsubsec:indirect_estimation}, Alice obtains $\pr_{key}$ and $\pr_{err}$, using the information received from Bob during the verification phase. This allows the parties to avoid the loss of key digits, at a price of higher uncertainty on the estimated values, which are calculated from several experimentally obtained quantities, each with its error. The results are reported in Table \ref{tab:table1}.

\begin{table}
\centering
\resizebox{\columnwidth}{!}{%
	\begin{tabular}{c|c|c|c}\hline\hline
	 &Direct Method\ &\ Direct Method\ &\ Indirect Method \\[0.8mm]
	& (full dataset) & (subset) & (full dataset) \\[0.8mm]
	\hline\hline
	$\pr_{key}$ & $1.55(3) \times 10^{-2}$ &  $1.5(1)\times 10^{-2}$ & $1.5(3)\times 10^{-2}$ \\[0.8mm]
	$\pr_{err}$ & $ 7.5(8) \times 10^{-4}$ & $5(2) \times 10^{-4}$ & $3(3)\times 10^ {-3}$ \\[0.8mm]
	\hline\hline
	\end{tabular}%
	}
	\caption{\footnotesize \textbf{Evaluation of key generation and error rates}. The probabilities of raw-key generation, $\pr_{key}$ and error on a key digit, $\pr_{err}$, respectively, are shown per round (in our case an interval of $0.5$ s). In the table, the numbers in parentheses are the errors on the last digits, obtained with the assumption of poissonian uncertainty on the counts.}	
	\label{tab:table1}
\end{table}

\begin{figure}[H]
\includegraphics[width=\linewidth]{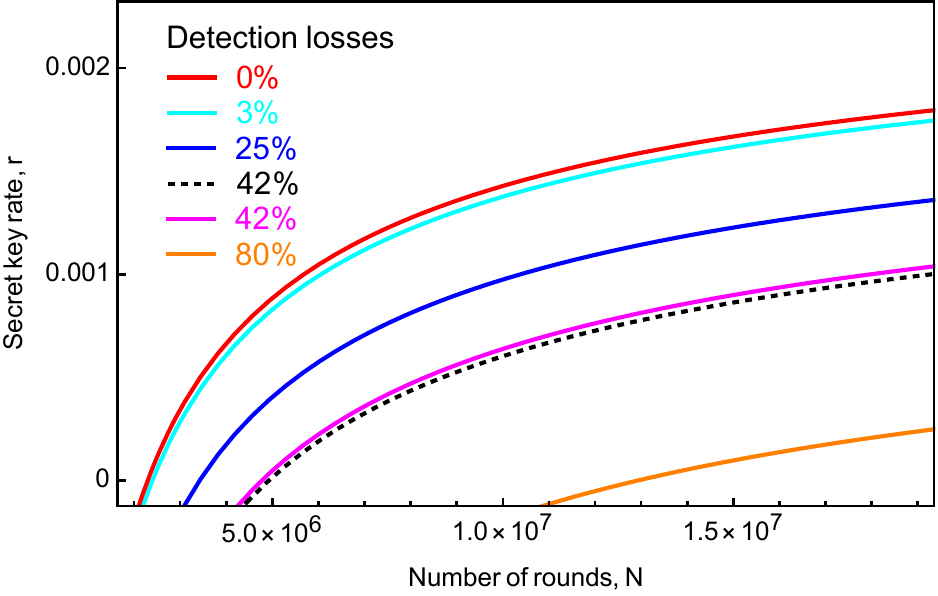}
\caption{\footnotesize \textbf{Secret key rate vs number of rounds, for different values of detection loss}. The black dashed curve refers to the experimental implementation, corresponding to a detection loss of $42\%$ for each Alice and Bob. The red, cyan, blue, magenta and orange curves represent the calculated results for  detection losses of $0,3,25,42$ and $80\%$, respectively. If the detection loss increases, the number of rounds for which $r$ becomes positive also increases, while the asymptotic secret key rate decreases. In the implemented case, the secret key rate becomes positive after about $4.9 \times 10^{6}$ rounds.}\label{fig:diff_eff_main}
\end{figure}

Based on the probabilities in Table~\ref{tab:table1}, we obtain the dependence of the final secret key rate, $r$, on the number of rounds, $N$, see Equation~\eqref{eq:secure_key-rate}. This dependence is plotted in Figure~\ref{fig:diff_eff_main}, for different values of the detection losses of D$_{\text{A}}$ and D$_{\text{B}}$, assumed to be the same. The details of how the curves were obtained are discussed in Appendices~ \ref{sec:bound_entropy} and~ \ref{sec:detector_eff}. As expected, an increase in the detection loss degrades the performance of the protocol.

\begin{figure}[H]
\includegraphics[width=\linewidth]{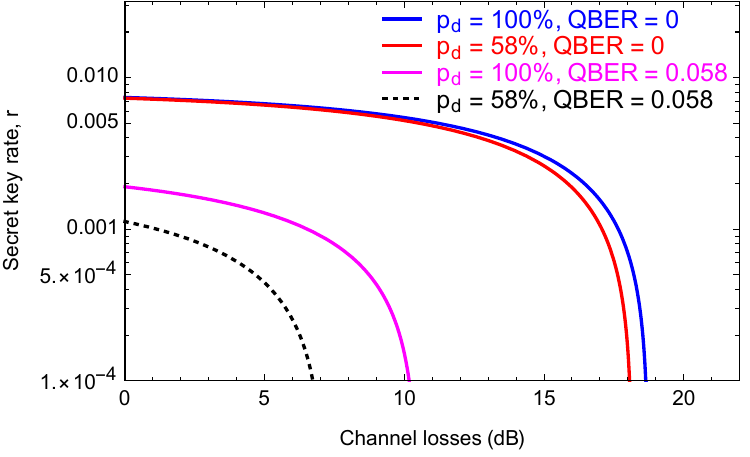}
\caption{\footnotesize \textbf{Secret key rate vs  channel transmission loss, for different values of detection efficiency and quantum bit error rate (QBER)}. The black dashed curve corresponds to the experimental parameters, with a detection efficiency of $58\%$ for each Alice and Bob and a QBER of $0.058$. The magenta, red and blue curve represent the three following ideal situations, respectively: perfect detectors and experimentally obtained QBER, imperfect detectors and QBER = 0, perfect detectors and QBER = 0. All curves are obtained by considering $10^9$ rounds of the protocol and the statistics of the used single-photon source.}
\label{fig:loss}
\end{figure}

We also report in Figure \ref{fig:loss}  the dependence of the secret key rate on the loss in the quantum channel between the server and each user, assumed to be the same for both, Alice and Bob. We present plots for different values of the detection efficiency and the quantum bit error rate (QBER), which is defined as the fraction of errors in the sifted key. More details on how these plots are obtained can be found in Appendix \ref{sec:losses}.

Given the results of Figure \ref{fig:loss}, we can compare the performance of our protocol to that of other QKD schemes. A natural candidate for the comparison is measurement-device-independent (MDI) QKD \cite{MDI-QKD}, which also involves an external server performing the detection. To our knowledge the best implementation to date of MDI-QKD achieves a secret key rate of about $10^{-4}$ for $7$ dB of channel transmission loss \cite{expMDI-QKD}. We obtain a similar key rate at the same transmission loss, as shown by the dashed line in Figure \ref{fig:loss}. However, the secret key rate for our experimental parameters quickly decreases for higher losses, contrary to the realization in \cite{expMDI-QKD}, where a secret key rate of $4.9 \times 10^{-6}$ is reported for $20.4$ dB of loss. Nevertheless, by considering QBER $= 0$, we obtain rates of the order of $10^{-4}$ for about $18$ dB of channel loss. These results indicate that our protocol can perform as good as MDI-QKD for transmission losses up to about $7$ dB, at least within the boundaries of our experimental implementation. At the moment, it is not clear if the performance could be made comparable also for higher losses, which however would require a more advanced experimental realization of our protocol.

Additionally, we stress that our estimated rates are lower-bounds and the actual key rates could be significantly higher. Indeed, to compute these lower bounds on $S(A|C)$, we took advantage of the strong sub-additivity of von Neumann entropy by actually discarding several components of the entropy function (components which would only have increased Eve's uncertainty -- thus, by discarding them, we are giving an unrealistic advantage to the adversary causing the key rate to drop). Such a method gives a worst-case computation.

\section{Conclusions}\label{sec:conclusions}

In our work, we propose and experimentally implement a novel QKD protocol allowing two classical users to establish a shared secret key using the services of an untrusted quantum server, which provides a superimposed single photon as a feasible quantum resource. We  underline the applicability of our scheme by providing an information-theoretic security analysis of our protocol in the finite-key setting, which takes into account imperfect detection efficiency and multi-photon emission from the source, and by calculating the secret key rate. 


Experimentally, the main challenge of the protocol is that it requires phase stability in the interferometer formed between the users and the server. This issue can be addressed by using intrinsically phase-stable schemes, like Sagnac configurations \cite{zhong2019}. In this case, however, a quantum channel between Alice and Bob is also necessary. 

As an immediate future line of research, our security analysis of finite keys in the presence of experimental imperfections can be applied to show the same security levels for other cryptographic schemes, such as counterfactual quantum cryptography \cite{noh:09, ren:11,bri:12,liu:12,cao:17}, or the key distribution based upon recently proposed two-way communication with one photon~\cite{san:dak:18, mas:18}.

In practical terms, recent progresses in bright deterministic single-photon sources~\cite{senellart2017}, high-efficiency detectors~\cite{dauler2014} and fast switches~\cite{lenzini2017} promise to push our scheme towards real-world applications.

\begin{acknowledgments}
We would like to thank Giulia Rubino for help with some figures and Borivoje Daki\'c and \"Amin Baumeler for useful discussions. P.Y., P.M., N.P. and A.S. acknowledge the support of SQIG -- Security and Quantum Information Group, the Instituto de Telecomunica\c{c}\~oes (IT) Research Unit, ref. UIDB/50008/2020 (actions QuRUNNER, QUESTS), funded by Funda\c{c}\~ao para a C\^{e}encia e Tecnologia (FCT), and the FCT projects QuantumMining POCI-01-0145-FEDER-031826, Predict PTDC/CCI-CIF/29877/2017 and  QuantumPrime PTDC/EEI-TEL/8017/2020, supported by the European Regional Development Fund (FEDER), through the Competitiveness and Internationalization Operational Programme (COMPETE 2020) of the Portugal 2020 framework [Project Q.DOT with Nr.\ 039728 (POCI-01-0247-FEDER-039728)], and by the Regional Operational Program of Lisbon. P.Y. acknowledges the support of DP-PMI and FCT (Portugal) through the scholarship PD/BD/113648/2015. W.K. is partially supported by NSF Grant No. 1812070. N.P.~acknowledges FCT project CERN/FIS-PAR/0023/2019, as well as the FCT Est\'{i}mulo ao Emprego Cient\'{i}fico grant no. CEECIND/04594/2017/CP1393/CT000. A.S. acknowledges funds granted to LaSIGE Research Unit, ref. UID/CEC/00408/2013. P.W. acknowledges support from the research platform TURIS, from the European Commission through ErBeStA (No.800942), from the Austrian Science Fund (FWF) through BeyondC (F7113-N48) and Research Group 5 (FG5), and from the U.S. Air Force Office of Scientific Research  (FA9550-21-1-0355) \textbf{Author contributions:} P.Y., W.K., P.M., N.P., A.S. developed the protocol and analysed its security. F.M., A.M. and P.W. designed and implemented the experiment and analysed the experimental data. All the authors contributed to the writing of the final manuscript. F.M. and P.Y. contributed equally to this work. \textbf{Corresponding authors:} correspondence to Francesco Massa (francesco.massa@univie.ac.at) or Philip Walther (philip.walther@univie.ac.at). 
\end{acknowledgments}

\onecolumngrid

\appendix

\section{Extraction of the secret key}\label{sec:key_extraction}
In order to compute the secret key rate described above, one needs to compute $S(A|C)$ for a given system. Before we proceed to discuss the ideal and experimental scenario, let us first define some useful terminology. 

Let us denote the Hilbert spaces corresponding to Alice's and Bob's equipments as $\mathcal H_A \!=\! \mbox{span}\{\ket{D_c}_{\!A}, \ket{D_v}_{\!A}, \ket{D_\ell}_{\!A}, \ket{D'_\ell}_{\!A}, \ket{D'_c}_{\!A}, \ket{R}_{\!A} \}$~and $\mathcal H_B \!=\! \mbox{span}\{\ket{D_c}_{\!B}, \ket{D_v}_{\!B}, \ket{D_\ell}_{\!B}, \ket{D'_\ell}_{\!B}, \ket{D'_c}_{\!B}, \ket{R}_{\!B} \}$,~respectively. Here, $\ket{D_c}$ and $\ket{D_v}$ denote the states of a detector, the first corresponding to the case of a photon causing a click, and the second corresponding to the case when there were no photons, resulting in a no-click. The detectors' state corresponding to the case when an incoming photon was lost is denoted as $\ket{D_\ell}$. The state $\ket{D'_\ell}$ corresponds to a loss, while $\ket{D'_c}$ to a click, of the photon at time $t' \neq t$, when two non-simultaneous photons were emitted by the source at times $t$ and $t'$. Finally, $\ket{R}$ denotes the state of a reflecting mirror. Note that the states corresponding to a click, $\ket{D_c}$ and $\ket{D'_c}$, and the ones corresponding to no-click, $\ket{D_v}$, $\ket{D_\ell}$ and $\ket{D'_\ell}$ are macroscopically distinguishable between each other as groups of those with or without clicks; and also to $\ket{R}$. However, the first two, $\ket{D_c}$ and $\ket{D'_c}$, are not distinguishable among each other, since in our set-up, Alice and Bob do not keep track of the detection times. Moreover, the latter three states, $\ket{D_v}$, $\ket{D_\ell}$ and $\ket{D'_\ell}$, also cannot be distinguished among each other, since without performing sophisticated quantum measurements, one cannot distinguish whether a detector did not click because there were no photons present, or they were lost.

We denote the server's Hilbert space as $\mathcal{H}_{S}=\mbox{span} \{ \ket 0_S ,\ket 1_S ,\ket v_S,\ket m_S\}$ consists of macroscopic orthogonal states modeling classical messages ``0'', ``1'', ``$v$'' (vacuum) and ``$m$'' (multiple clicks), respectively. Additionally, we denote server's ancilla system by $C$, spanned by the Hilbert space $\mathcal H_C$, which a dishonest server can entangle with the photons sent to Alice and Bob to extract information about the exchanged key.

Let us assume Alice tosses a fair coin to decide whether she will detect or reflect the photon, and set the initial state of the apparatus accordingly, resulting in a proper mixture of the two states, $\ket{D_v}_{\! A} \bra{D_v}$ and $\ket{R}_{\! A} \bra{R}$, and analogously for Bob. Without the loss of generality, we can always include the coin states into the macroscopic description of the apparatus states, such that the purified initial state of Alice's apparatus is
\begin{equation}
\ket{\phi_0}_{\!A}=\dfrac{1}{\sqrt 2} \Big( \ket{D_v}_A + \ket R_A \Big),
\end{equation}
and analogously for Bob, making their joint state as
\begin{equation}\label{eq:AB}
\ket{\phi_0}_{\!AB}\!=\!\frac 1 2 \!\ \!\Big(\!\ \!\!\! \ \ket{D_v,\! R}_{\!AB} + \ket{R,\! D_v}_{\!AB} + \ket{D_v,\! D_v}_{\!AB} + \ket{R,\! R}_{\!AB} \!\!\Big).
\end{equation}

Note that due to possible imperfect single-photon sources, and the presence of adversaries, the number of photons present is not necessarily fixed to be one. Thus, we will use a number basis to describe the photonic states. In this paper, we will decompose the overall Fock space of the photons in Alice's and Bob's arms as $\mathcal F_f \!=\! \mbox{span}\{ \ket{0,0}_{\!f}, \ket{1,0}_{\!f}, \ket{0,1}_{\!f}, \ket{2,0}_{\!f}, \ket{1,1'}_{\!f}, \ket{1',1}_{\!f}, \ket{0,2}_{\!f}\!\}$ $\oplus \mathcal{F}^k_f$, where $\ket{0,0}_f \equiv \ket v_f$ represents the vacuum state, $\ket{1,0}_f$ represents a photon in Alice's arm and $\ket{0,1}_f$ to be in Bob's arm. Similarly, $\ket{2,0}_f,$ and $\ket{0,2}_f,$ represent two non-simultaneous photons in Alice' and Bob's arms, respectively; whereas $\ket{1,1'}_f$ and $\ket{1',1}_f$ represent the case of two non-simultaneous photons when the first one went to Alice's arm while the second to Bob and vice-versa, respectively. $\mathcal{F}^k_f$ denotes the sub-space corresponding to the multi-photon case of $k>2$ photons. The action of photonic  creation operators $\hat{a}^{\dagger}$ and $\hat{b}^\dagger$, in terms of the number basis $\ket{a,b}_f$, with $a,b \in \mathbb{N}_0$ being the number of photons in Alice's and Bob's arms, respectively, is given by $(\hat{a}^\dagger)^a (\hat{b}^\dagger)^b \ket{0}_f = \sqrt{a! \ \! b!}\ket{a, b}_f$. We can now proceed to analyze the experimental implementation of our protocol with imperfect single-photon sources and detectors, as well as the noisy and lossy channels.

We assume an untrusted server that can attack before Alice and Bob perform their respective operations, as well as after (which is equivalent to allowing Eve to intercept the photons exchanged between an honest server and the agents). We consider a poissonian probabilistic single photon source, emitting vacuum state with probability $\pr_0$, single photons with probability $\pr_1$, two non-simultaneous photons with probability $\pr_2$, etc., within a time slot of interval $T$,~as
\begin{eqnarray}
\!\!\!\ket{\phi_0}_f&=&\sqrt{\pr_0} \ket{v}_f + \sqrt{\dfrac{\pr_1}{T}} \int_0^T \hat{a}^\dagger(t) \ket v_f \mbox{d}t + \dfrac{\sqrt{\pr_2}}{T} \! \int_0^T \!\! \int_0^T \!\! \Bigg(\!\dfrac{\hat{a}^\dagger(t) \hat{a}^\dagger(t')}{\sqrt{2}} \ket v_f \!\Bigg) \mbox{d}t \!\mbox{ d}t' \!+ \dots ,
\end{eqnarray}
where $\hat{a}^\dagger(t)$ and $\hat{a}^\dagger(t')$ represent photon creation at times $t$ and $t'$, respectively. In our particular implementation, the average number of photons is $0.35$, yielding $\pr_0 = 0.705$, $\pr_1 = 0.247$, $\pr_2 = 0.043$. For simplicity, and in order to compare the theoretical analysis with our experimental data, the probability to emit higher numbers of photons is considered negligible, i.e., $\pr_0 + \pr_1 + \pr_2 \approx 1$. Thus, the initial photon state is
\begin{eqnarray}
\ket{\phi_0}_{f} &=&\sqrt{\pr_0} \ket{v}_f + \sqrt{\pr_1} \ket 1_f + \sqrt{\pr_{2}} \ket{2}_f,
\end{eqnarray}
where $\ket{v}_f  \equiv \ket{0}_f$ is the photon vacuum state, $\ket 1_f = \hat{a}^\dagger(t) \ket v_f$, $\sqrt{2}\ket 2_f = \hat{a}^\dagger(t) \hat{a}^\dagger(t') \ket v_f$. Nevertheless, our analysis can straightforwardly generalised to an arbitrary number of emitted photons. Note that, for simplicity, we omitted the time integrals in the definition of the single- and two-photon states, $\ket 1_f$ and $\ket 2_f$, respectively, as we consider that the users do not keep track of the photon detection times, meaning that, at the end of each round, Alice, Bob and the server only have access to the number of detections they recorded. This makes our analysis also applicable to the case of simultaneous multi-photon emission.

After passing through the first 50/50 beam splitter of our interferometer, described by $\hat{a}^\dagger(t) \! \rightarrow\! (\hat{a}^\dagger(t) + \hat{b}^\dagger(t))/\sqrt 2$ and $\hat{a}^\dagger(t') \! \rightarrow \!(\hat{a}^\dagger(t') + \hat{b}^\dagger(t'))/\sqrt 2$, the above state becomes
\begin{eqnarray}\label{eq:41}
\!\!\!\!\ket{\phi_0}_{\!f} &=&\sqrt{\pr_0} \!\; \ket{v}_{\!f} + \sqrt{\frac{\pr_1}{2}} \Big(\ket{1,0}_{\!f} + \ket{0,1}_{\!f} \Big) + \frac{\sqrt{\pr_{2}}}{2} \Big( \ket{2,0}_{\!f} + \ket{1,1'}_{\!f} + \ket{1',1}_{\!f} + \ket{0,2}_{\!f} \Big).
\end{eqnarray}
Upon possible further action of the adversary, the most general photon-server (normalized) state is given by
\begin{eqnarray}\label{eq:initial_photon_state}
\ket{\phi_0}_{\!fC} \!&=& \!\!\!\sum_{\substack{a,b \;\!\geq\;\! 0\\a+b\;\!\leq\;\!2}} \!\!\ket{a,b}_{\!f} \ket{c_{a,b}}_{C} \nonumber\\
&=&\! \ket{0,0}_{\!f} \!\otimes \ket{c_{0,0}}_{\!C} \nonumber \\[0.8mm] &&+ \ket{1,0}_{\!f} \!\otimes \ket{c_{1,0}}_{\!C} \!+ \ket{0,1}_{\!f} \!\otimes \ket{c_{0,1}}_{\!C}\nonumber \\[0.8mm]
&& + \!\; \ket{2,0}_{\!f} \!\otimes \ket{c_{2,0}}_{\!C} + \ket{0,2}_{\!f} \!\otimes \ket{c_{0,2}}_{\!C} + \!\; \ket{1,1'}_{\!f} \!\otimes \ket{c_{1,1'}}_{\!C} + \ket{1',1}_{\!f} \!\otimes \ket{c_{1',1}}_{\!C}.
\end{eqnarray}
where $\ket{c_{a,b}}_C \in \mathcal{H}_C$ (not necessarily orthogonal, nor normalized states) are associated to the cases when there are $a$ and $b$ photons entering Alice's and Bob's arms, respectively. Nevertheless, the states $\ket{c_{a,b}}_C$ are arbitrary and contain any number of photons. Therefore, the overall state before the photon(s) enter Alice's and Bob's labs is
%
\begin{eqnarray}\label{eq:initial_state}
\ket{\phi_0}_{\!AB\!fC} \!&=& \ket{\phi_0}_{AB} \!\otimes \ket{\phi_0}_{fC} \nonumber \\
&=&\! \frac 1 2 \!\Big(\!\!\ket{D_v,\! D_v}_{\!\!AB} \!+\! \ket{D_v,\! R}_{\!\!AB} \!+\! \ket{R,\! D_v}_{\!\!AB} \!+\! \ket{R,\! R}_{\!\!AB} \!\!\Big)\! \nonumber\\
&& \otimes \Big(\!\ket{0,0}_{\!f} \otimes \ket{c_{0,0}}_{\!C} \nonumber\\
&& \quad \ + \!\; \ket{1,0}_{\!f} \otimes \ket{c_{1,0}}_{\!C} \!+ \ket{0,1}_{\!f} \otimes \ket{c_{0,1}}_{\!C} \nonumber  \\
&& \quad \ + \!\; \ket{2,0}_{\!f} \otimes \ket{c_{2,0}}_{\!C} + \ket{0,2}_{\!f} \otimes \ket{c_{0,2}}_{\!C} + \!\; \ket{1,1'}_{\!f} \otimes \ket{c_{1,1'}}_{\!C} + \ket{1',1}_{\!f} \otimes \ket{c_{1',1}}_{\!C} \!\!\Big).
\end{eqnarray}

Let us denote Alice's and Bob's respective detectors' efficiencies as $\pr_d^A$ and $\pr_d^B$, with the respective losses being $\pr_\ell^A=1-\pr_d^A$ and $\pr_\ell^B=1-\pr_d^B$. In our experimental implementation, the two efficiencies are almost the same, with $\pr_d^A \approx \pr_d^B \approx 58\%$. The individual actions of, say, Alice, in this practical scenario are
\begin{equation}\label{eq:actions_exp}
\begin{array}{ll}
\!\!\!\!\!\!\!\ket{D_v}_{\!A} \ket{0}_{\!f} \rightarrow \ket{D_v}_{\!A} \ket{0}_{\!f},\\[1.6mm]
\!\!\!\!\!\!\!\ket{D_v}_{\!A} \ket{1}_{\!f} \rightarrow \left(\sqrt{\pr_\ell^A} \ket{D_\ell}_{\!A} + \sqrt{\pr_d^A} \ket{D_c}_{\!A} \right) \!\ket{0}_{\!f},\\[2.2mm]
\!\!\!\!\!\!\!\ket{D_v}_{\!A} \ket{2}_{\!f} \rightarrow \Big(\pr_\ell^A \ket{D_\ell D'_\ell}_{\!A} + \sqrt{\pr_\ell^A \pr_d^A}\ket{D_c D'_\ell}_{\!A} \\[1.5mm]
\!\!\!\!\!\quad\quad\quad\quad\quad \ + \sqrt{\pr_\ell^A \pr_d^A} \ket{D_\ell D'_c}_{\!A} + \pr_d^A \ket{D_c D'_c}_{\!A}\!\Big)\! \ket{0}_{\!f}, \\[4.5mm]
\!\!\!\!\!\!\!\ket R_{\!A} \ket 0_{\!f} \rightarrow \ket R_{\!A} \ket 0_{\!f},\\[1.5mm]
\!\!\!\!\!\!\!\ket R_{\!A} \ket 1_{\!f} \rightarrow \ket R_{\!A} \ket 1_{\!f}, \\[1.5mm]
\!\!\!\!\!\!\!\ket R_{\!A} \ket 2_{\!f} \rightarrow \ket R_{\!A} \ket 2_{\!f},
\end{array}
\end{equation}
where primed and unprimed states of the apparatuses correspond to at times $t'$ and $t$, respectively. Note that we assume that Alice and Bob trust their detectors with their finite detection efficiencies. Therefore, upon applying $U_1$, given in terms of Alice's and Bob's local actions described by~\eqref{eq:actions_exp}, we obtain the state $\ket{\phi_1}_{ABfC}=U_1 \ket{\phi_0}_{ABfC}$.

Upon leaving Alice's and Bob's labs, the server (or Eve) will apply a quantum instrument to the returning photon-server state. This can be modelled as an isometry $\mathcal{I}: \mathcal{F}_f \otimes \mathcal{H}_{C} \rightarrow \mathcal{H}_S \otimes \mathcal{H}_{C}$, given by 
\begin{eqnarray}\label{eq:I}
\!\!\!\!\!\!\!\!\!\!\mathcal{I}\ket{a',b'}_{\!f} \ket{c_{a,b}}_{\!C}&=& \ket{0}_S \ket{e_{a',b'}^{a,b}}_{\!C} + \ket{1}_S\ket{f_{a',b'}^{a,b}}_{\!C} + \ket{v}_S\ket{g_{a',b'}^{a,b}}_{\!C} + \ket{m}_S\ket{h_{a',b'}^{a,b}}_{\!C},
\end{eqnarray}
where states $\ket{e_{a',b'}^{a,b}}_{C}, \ket{f_{a',b'}^{a,b}}_{C}, \ket{g_{a',b'}^{a,b}}_{C}, \ket{h_{a',b'}^{a,b}}_{C} \in \mathcal{H}_C$ are again not necessarily normalized, nor orthogonal. Note that, due to the action of $U_1$, the photon numbers $a,b$ are no longer correlated to $a',b' \in \{ 0,1, 2 \}$; nevertheless, we still have $a' + b' \leq 2$. From this, one obtains the final state between the users and the server, $\ket{\phi_2}_{ABSC}=\mathcal I \ket{\phi_1}_{ABfC}$. Using Theorem \ref{thm:cq-entropy}, one can lower bound the conditional entropy $S(A|C)$, as explained in detail in the next section.

\section{Two particular attacks}\label{sec:muçti-photon}

\subsection{Adaptive attack with a single photon}\label{subsec:bomb}

\begin{figure}[H]
  \centering
  \includegraphics[width=0.4\linewidth]{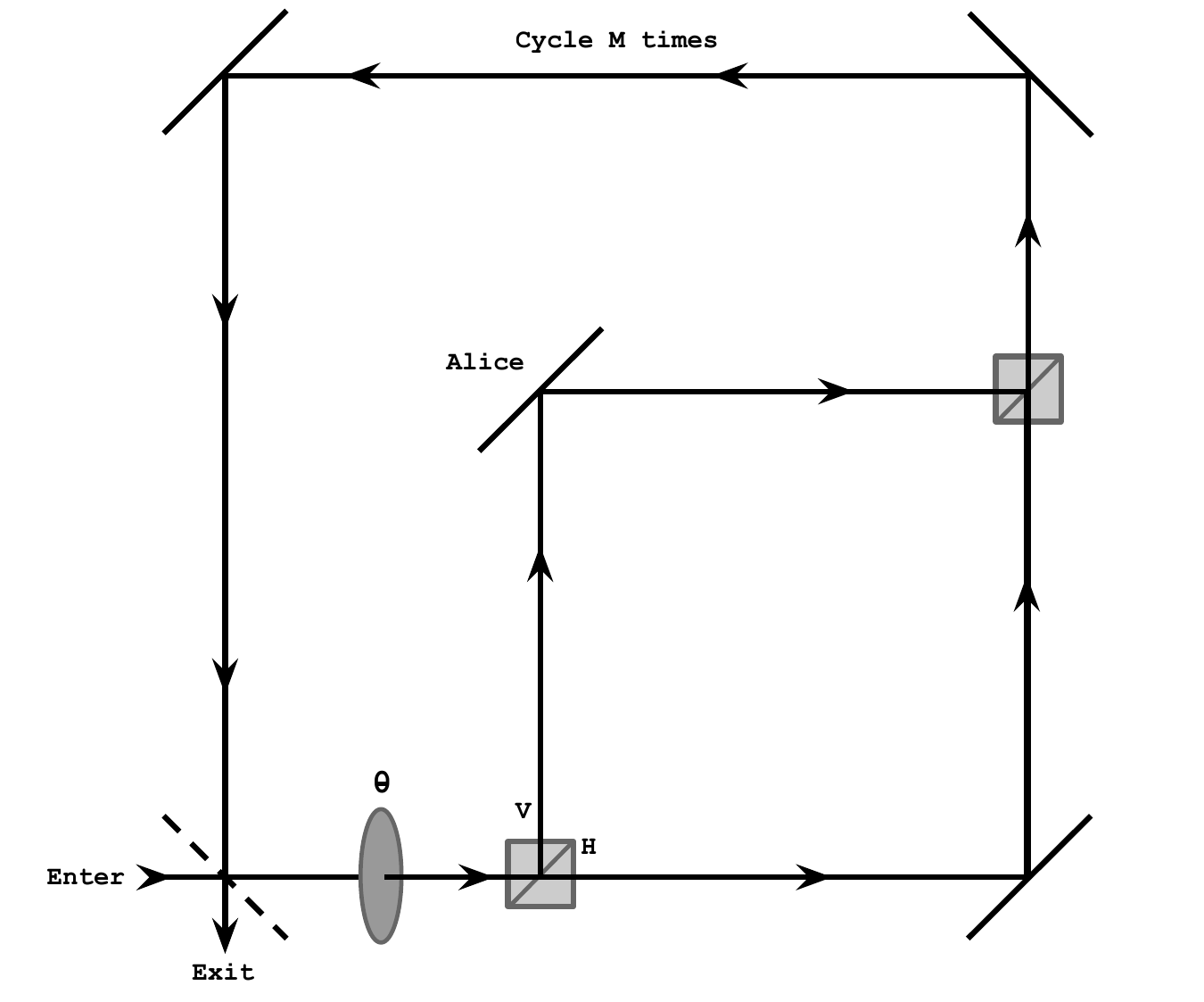}
\caption{\footnotesize \textbf{Eve's attack based on interaction-free measurement.} }\label{fig:bomb}
\end{figure}

The adaptive attack with a single photon that is fed in an agent's laboratory several times during a single round of the key distribution protocol is based on the interaction-free measurement proposed in~\cite{rud:gro:02}, depict in Figure~\ref{fig:bomb}. An agent, say Alice, is placed in one arm of an interferometer which consist of an input polarizing beam splitter and standard balanced beam splitter on its output. Before entering the interferometer, the initial polarization state, say horizontal state $\ket{\psi_0} = \ket{H}$, is rotated by a certain angle $\theta$, so that before the polarizing beam splitter it is $\ket{\psi_\theta} = \cos\theta \ket{H} + \sin\theta \ket{V}$. In case Alice decided to ``reflect'', at the output of the interferometer the polarization state of the photon will stay the same, $\ket{\psi_\theta}$. In case she decided to ``detect'', with probability $\sin^2\theta$ the photon will end up in Alice's laboratory and be absorbed, while with probability $\cos^2\theta$ it will leave the interferometer in polarization state $\ket{\psi_0}$. In the case of the latter, the process is repeated, up to $M$ times. If the rotation angle is chosen to be $\theta = \pi/2M$, after $M$ iterations the polarization state will be $\ket {\psi_{\pi/2}} = \ket V$ in case Alice decided to ``reflect'', while it will stay ``frozen'' to $\ket {\psi_{0}} = \ket H$ in case she decided to ``detect'', i.e., the two states will be fully distinguishable, and Eve would know Alice's action. The probability that a photon will not end in Alice's arm $M$ consecutive times when she decided to ``detect'' is $p = \cos^{2M}\theta = (\cos\frac{\pi}{2M})^{2M}$, which for large $M$ behaves like $p \sim 1 - \pi^2/4M \rightarrow 1$. Thus, with probability arbitrarily close to 1 Eve can learn Alice's action without triggering her detector (``activating the bomb'' from the original scenario discussed in~\cite{rud:gro:02}).

But in our case, the aim of Eve is to simulate, as much as possible, the honest scenario, in which Alice's detector will click in about half of the cases. This is achieved with pretty good accuracy for $M=4$ already, as we have that $p = (\cos\frac{\pi}{8})^8 \approx 0.53$. Nevertheless, if Eve wanted to learn the actions of both agents, she would need to perform two such measurements performed on both agents. But this would inevitably lead to increased double clicks in rounds when both agents decide to ``detect'' (Note that in order to learn the action of a single agent, say Alice, Eve should perform measurement after her laboratory, thus destroying any possible coherence between photon(s) state in Alice's and Bob's labs). During the parameter estimation phase, Alice and Bob can infer such increased probability of coincidences, and thus detect eavesdropping.

\subsection{Multi-photon attack}\label{subsec:noon}

This is a version of the above interaction attack in which instead of sending a single photon through the interferometer $M$ times, Eve sends $M$ photons only once, in order to learn action of a single agent. Thus, it suffers from the same deficiency as the previous attack: Alice's photon detection is not correlated with Bob's one and therefore will change the joint detection statistics. Again, note that in order to learn the action of a single agent, Eve must perform her measurement on the photons outside her/his lab, thus destroying any possible coherence. In other words, sending a coherent superposition between photon states sent to Alice and Bob offers no advantage.

But this attack features additional problem, in that Eve cannot fully distinguish between an agent's actions, leading her to announce inconsistent messages allowing  Alice and Bob to additionally detect cheating. Let us first describe this attack in more detail. Eve sends a multi-photon state $\ket{\Psi_\theta(M)} = \ket{\psi_\theta}^{\otimes M} = (\cos\theta \ket{H} + \sin\theta \ket{V})^{\otimes M}$. If Alice decides to ``reflect'', Eve will receive the same $M$-photon state $\ket{\Psi_\theta(M)}$ at the output of the interferometer. In case she decides to ``detect'' and at least one of the photons ends in her arm, there will be less then $M$ photons at the output of the interferometer, and Eve can thus infer Alice's action. But if not a single photon gets detected by Alice, at the output of the interferometer we would have the $M$-photon state $\ket{\Psi_0(M)} = \ket{\psi_0}^{\otimes M} = \ket{H}^{\otimes M}$. Thus, Eve cannot distinguish the two actions by measuring the photon number, and she needs to subsequently perform polarization measurement. The optimal discrimination probability for the two states is given in terms of the transition probability $\tilde p = |\langle\Psi_0 | \Psi_\theta \rangle|^2 = |\langle\psi_0 | \psi_\theta\rangle|^{2M} = \cos^{2M}\theta$, which is precisely the probability that in the case of deciding to ``detect'' none of $M$ photons end up in Alice's arm. On the other hand, as before we want that this probability is equal to 1/2, to match the honest scenario. Thus, if Eve wants to emulate the honest scenario, she must set $\theta$ such that the output polarization states are far from fully distinguishable. In other words, the adversary will necessarily occasionally announce messages that are inconsistent with the agents' actions, thus revealing eavesdropping. One can straightforwardly apply our methodology to this case to obtain quantitative expression for the secret key rate. Therefore, we omit this rather complex, but straightforward analysis.


\section{Security Analysis - Ideal case}\label{sec:ideal}
\red{In Appendix \ref{sec:bound_entropy}, we show how to prove the security of our protocol in the general case, assuming practical devices.  To develop the intuition behind the proof in that section, however, we first consider the ideal case scenario.  Here}, we assume that the server has a perfect single-photon source, Alice's and Bob's detectors are perfect, which means they have $100 \%$ detection efficiency and zero dark counts, but there may be channel loss.
Therefore, the perfect single photon state that Alice and Bob expect to be sent is
\begin{equation}
\ket{\phi_0}_f = \left(\dfrac{\hat{a}^\dagger + \hat{b}^\dagger}{\sqrt 2}\right) \ket{0,0}_f = \dfrac{\ket{1,0}_f + \ket{0,1}_f}{\sqrt 2},
\end{equation}
with $\ket{1,0}_f$ and $\ket{0,1}_f$ representing the photon located in Alice's and Bob's arms, respectively. However, we assume that the following entangled state is sent to Alice and Bob by the server (or Eve)
\begin{equation}
\ket{\phi_0}_{fC}= \ket{0,0}_f\otimes\ket{c_{0,0}} + \ket{1,0}_f \otimes \ket{c_{1,0}}_C + \ket{0,1}_f \otimes \ket{c_{0,1}}_C
\end{equation}
where $\ket{c_{a,b}}_C \in \mathcal{H}_C$ are not necessarily orthogonal nor normalized. Note that, this is the state arriving at $A$ and $B$'s lab, and so it also incorporates channel loss in the $\ket{0,0}_f\otimes\ket{c_{0,0}}$ term.  Moreover, as per usual in QKD security proofs, Alice and Bob can enforce symmetry, and so, we may assume $\bk{c_{0,1}}_C=\bk{c_{1,0}}$. Therefore, we can write the joint initial state as
\begin{eqnarray}
\!\!\!\!\ket{\phi_0}_{ABfC} &=& \ket{\phi_0}_{AB} \otimes \ket{\phi_0}_{fC} \\[0.5mm]
\!\!\!\!&=& \dfrac 1 2 \Big(\! \ket{D_v,\! R}_{\!AB} + \ket{R,\! D_v}_{\!AB} + \ket{D_v,\! D_v}_{\!AB} + \ket{R,\! R}_{\!AB}\! \Big)\!\nonumber\\
&& \otimes \!\left(\ket{0,0}_f\ket{c_{0,0}}_C + \ket{1,0}_f \ket{c_{1,0}}_C + \ket{0,1}_f \ket{c_{0,1}}_C \right). \nonumber
\end{eqnarray}
Alice's and Bob's actions on a given initial photon state are given by
\begin{equation}
\begin{array}{ll}\label{eq:actions_ideal}
\ket{D_v,\! R} \ket{1,0} \rightarrow \ket{D_c,\! R} \ket{0,0},\quad \quad\quad \quad \quad \quad& \ket{R,\! D_v} \ket{1,0} \rightarrow \ket{R,\! D_v} \ket{1,0}, \\[0.8mm]
\ket{D_v,\! R} \ket{0,1} \rightarrow \ket{D_v,\! R} \ket{0,1},\quad \quad\quad \quad \quad \quad& \ket{R,\! D_v} \ket{0,1} \rightarrow \ket{R,\! D_c} \ket{0,0},  \\[1.5mm]
\ket{D_v,\! D_v} \ket{1,0} \rightarrow \ket{D_c,\! D_v} \ket{0,0},\quad \quad\quad \quad \quad \quad& \ket{R,\! R} \ket{1,0} \rightarrow \ket{R,\! R} \ket{1,0}, \\[0.8mm]
\ket{D_v,\! D_v} \ket{0,1} \rightarrow \ket{D_v,\! D_c} \ket{0,0},\quad \quad\quad \quad \quad \quad& \ket{R,\! R} \ket{0,1} \rightarrow \ket{R,\! R} \ket{0,1}, \\[0.8mm]
\ket{D_v,\! D_v} \ket{0,0} \rightarrow \ket{D_v,\! D_v} \ket{0,0},\quad \quad\quad \quad \quad \quad& \ket{R,\! R} \ket{0,0} \rightarrow \ket{R,\! R} \ket{0,0}, \\[0.8mm]
\ket{D_v,\! R} \ket{0,0} \rightarrow \ket{D_v,\! R} \ket{0,0},\quad \quad\quad \quad \quad \quad& \ket{R,\! D_v} \ket{0,0} \rightarrow \ket{R,\! D_v} \ket{0,0},
\end{array}
\end{equation}
and, therefore
\begin{eqnarray}
\!\!\!\!\!\ket{\phi_1}_{ABfC} = \dfrac{1}{2} &&\Big[\!\ket{D_c,\! R} \ket{0,0} \ket{c_{1,0}} + \ket{D_v,\! R} \ket{0,1} \ket{c_{0,1}} + \ket{R,\! D_v} \ket{1,0} \ket{c_{1,0}}_C + \ket{R,\! D_c} \ket{0,0} \ket{c_{0,1}} \nonumber \\
&& +  \ket{D_c,\! D_v} \ket{0,0} \ket{c_{1,0}} + \ket{D_v,\! D_c} \ket{0,0} \ket{c_{0,1}} + \ket{R,\! R} \!\big(\! \ket{1,0} \ket{c_{1,0}} + \ket{0,1} \ket{c_{0,1}}\\
&&+ \ket{D_v,\! D_v} \ket{0,0}\ket{c_{0,0}} + \ket{D_v,\! R} \ket{0,0}\ket{c_{0,0}} + \ket{R,\! D_v} \ket{0,0}\ket{c_{0,0}} + \ket{R,\! R} \ket{0,0}\ket{c_{0,0}} \Big].\nonumber
\end{eqnarray}

Following this, as in the experimental case, the adversary will apply a quantum instrument to the returning photon state which, as before, can be modeled as an isometry, whose action is defined as 
\begin{equation}\label{eq:I_ideal}
\mathcal{I}\ket{a',b'}_f \ket{c_{a,b}}_{C}= \ket{0}_S \ket{e_{a',b'}^{a,b}}_{C} + \ket{1}_S\ket{f_{a',b'}^{a,b}}_{C} + \ket{v}_S\ket{g_{a',b'}^{a,b}}_{C},
\end{equation}
where states from $\mathcal{H}_C$ are not necessarily normalized nor orthogonal, and $a,b$ are no longer correlated with $a',b'$ due to Alice's and Bob's actions given by Equation~\eqref{eq:actions_ideal}. Note that since we are assuming an  ideal case, the term corresponding to the message ``$m$'' is absent from the above equation.

We are interested only in the rounds when the server announces ``1'' and neither Alice nor Bob detect a photon, and the users generate the key. Thus, while writing the state after the server applies $\mathcal I$ on $\ket{\phi_1}_{ABfC}$, we will omit writing the server's message state $\ket{1}_S$ (corresponding to announcing a result ``1''). The final density operator representing the state of the system $ABC$, conditioned on the event that the server sends the message ``1'' and none of the users detects a photon (only the rounds used for key generation), is
\begin{eqnarray}\label{eq:phi_final2_ideal}
\ket{\phi_2}_{\!ABC} &&= \dfrac{1}{\sqrt\mathcal{N}} \;\!\! \Big\{\! \ket{D_v,\! R} \;\!\! \otimes  \;\!\! \ket{k_{0,0}} + \ket{R ,\!D_v} \;\!\! \otimes  \;\!\! \ket{k_{1,1}} + \ket{R,\!R} \;\!\! \otimes  \;\!\! \ket{k_{1,0}}  + \ket{D_v,\!D_v} \;\!\! \otimes  \;\!\! \ket{k_{0,1}} \! \Big\},
\end{eqnarray}
where the states $\ket{k_{i,j}}_{C}$ are associated to Alice establishing the value $i$ and Bob $j$ as a key bit, are given by
\begin{equation}
\begin{array}{ll}\label{eq:kijideal}
&\ket{k_{0,0}}_C = \dfrac{1}{2} \left[\ket{f_{0,1}^{0,1}} + \ket{f_{0,0}^{0,0}}\right], \\[4mm]
&\ket{k_{1,1}}_C = \dfrac{1}{2} \left[\ket{f_{1,0}^{1,0}} + \ket{f_{0,0}^{0,0}}\right], \\[4mm]
&\ket{k_{0,1}}_C = \dfrac{1}{2} \ket{f_{0,0}^{0,0}}, \\[4mm]
&\ket{k_{1,0}}_C = \dfrac{1}{2} \left[\ket{f_{1,0}^{1,0}} + \ket{f_{0,1}^{0,1}} + \ket{f_{0,0}^{0,0}}\right].
\end{array}
\end{equation}
Note that, though we are assuming in this ideal setting, that $A$ and $B$'s devices are ideal, the adversarial server may still ``simulate'' imperfect detectors which may have, for instance, dark counts (incorporated in the term $\bk{f_{0,0}^{0,0}}$ which is the probability the server sends a positive message in the event a vacuum actually enters its lab). The normalization constant $\mathcal{N}$ is, again, the probability to obtain the result $1$, $\pr(1)$, when there were no clicks at the users' detectors, and is given by
\begin{eqnarray}
&&\mathcal{N} = \bk{k_{0,0}} + \bk{k_{1,1}} + \bk{k_{1,0}} + \bk{k_{0,1}}= \pr (1).
\end{eqnarray}
As in the experimental case, we define $\pr_{0,0}=\pr(D_v,\! R \ ; 1)=\bk{k_{0,0}}$ as the joint probability for the event when Alice detects vacuum and Bob reflects, and the server announces the result ``1'', and analogously $\pr_{1,1}$, $\pr_{0,1}$ and $\pr_{1,0}$. Again, we use the semicolon (;) to denote logical AND operation between two propositions. Therefore, we can define the probability to share the key as $\pr_{key}=\pr_{0,0}+\pr_{1,1}$ and the probability of an error as $\pr_{err}=\pr_{0,1}+\pr_{1,0}$. When we evaluate our key rate bound, we use $\mathcal Q$ to be the probability that the server announces the result ``1'', given both Alice and Bob reflected, conditioned on a photon arriving at the server. Finally, we allow the adversarial server to ``simulate'' dark counts at a rate of $p_d$ (to its advantage), and we use $T$ to mean the probability of transmittance in one direction, namely $1-T$ is the probability the photon is dropped before it gets to $A$ or $B$ (the probability the photon returns to the server if $A$ and $B$ reflect is $T^2$).  In the ideal case, it is easy to see that
\begin{equation}
\begin{array}{ll}\label{eq:theoretic_values}
\!\!\!\!\pr_{0,0}=\bk{k_{0,0}} = \dfrac{1}{4}\left(\dfrac{T^2}{4} + \dfrac{T(1-T)p_d}{2}\right) \ ,\ \ & \pr_{0,1}=\bk{k_{0,1}} = \dfrac{(1-T)p_d}{4} \ \ , \\[3mm]
\!\!\!\!\pr_{1,1}=\bk{k_{1,1}} = \dfrac{1}{4}\left(\dfrac{T^2}{4} + \dfrac{T(1-T)p_d}{2}\right) \ ,\ \ & \pr_{1,0}=\bk{k_{1,0}} =  \dfrac{\mathcal Q\cdot T^2}{4} \ .
\end{array}
\end{equation}

Expanding $\mbox{Re}\braket{k_{0,0}|k_{1,1}}$, needed for the entropy bound computation, we find:
\begin{equation}\label{eq:newk00}
\mbox{Re}\braket{k_{0,0}|k_{1,1}} = \frac{1}{4}(\mbox{Re}\braket{f_{0,1}^{0,1}|f_{1,0}^{1,0}} + \mbox{Re}\braket{f_{0,0}^{0,0}|f_{0,0}^{0,0}} + \mbox{Re}\braket{f_{0,1}^{0,1}|f_{0,0}^{0,0}} + \mbox{Re}\braket{f_{0,0}^{0,0}|f_{1,0}^{1,0}})
\end{equation}
Expanding $\braket{k_{1,0}|k_{1,0}}$ we find:
\begin{align}
&\braket{k_{1,0}|k_{1,0}} = \frac{1}{4}(\bk{f_{1,0}^{1,0}} + \bk{f_{0,1}^{0,1}} + \bk{f_{0,0}^{0,0}} + 2\mbox{Re}(\braket{f_{0,1}^{0,1}|f_{1,0}^{1,0}} + \braket{f_{0,1}^{0,1}|f_{0,0}^{0,0}} +  \braket{f_{1,0}^{1,0}|f_{0,0}^{0,0}})).\notag\\
\Rightarrow& \mbox{Re}(\braket{f_{0,1}^{0,1}|f_{1,0}^{1,0}} + \braket{f_{0,1}^{0,1}|f_{0,0}^{0,0}} +  \braket{f_{1,0}^{1,0}|f_{0,0}^{0,0}}) = \frac{4\bk{k_{1,0}} - (\bk{f_{1,0}^{1,0}} + \bk{f_{0,1}^{0,1}} + \bk{f_{0,0}^{0,0}})}{2}\label{eq:newk10}
\end{align}
Substituting Equation~\eqref{eq:newk10} into Equation~\eqref{eq:newk00} we have:
\begin{align}
\mbox{Re}\braket{k_{0,0}|k_{1,1}} = \frac{1}{2}\bk{k_{1,0}} - \frac{1}{8}(\bk{f_{1,0}^{1,0}} + \bk{f_{0,1}^{0,1}} + \bk{f_{0,0}^{0,0}}).
\label{eq:ideal_overlap}
\end{align}
The term $\bk{f_{0,0}^{0,0}}$ is an observable quantity, it is simply $4\bk{k_{0,1}} = 4p_{0,1}$.  The values of $\bk{f_{0,1}^{0,1}}$ and $\bk{f_{1,0}^{1,0}}$ can be bounded by solving the following quadratic equation (derived from the expansion of $\bk{k_{0,0}}$ and $\bk{k_{1,1}}$ respectively):
\begin{equation}
\bk{f_{0,1}^{0,1}} + 2\cos\theta\sqrt{\bk{f_{0,0}^{0,0}}} \sqrt{\bk{f_{0,1}^{0,1}}} + (\bk{f_{0,0}^{0,0}} - 4\bk{k_{0,0}}).
\end{equation}
Similarly for $\bk{f_{1,0}^{1,0}}$.  This allows us to minimize $|\mbox{Re}\braket{k_{0,0}|k_{1,1}}|$, thus minimizing the adversary's uncertainty (i.e., minimizing $S(A|C)$).

At this point, we compute the conditional entropy between Alice and the adversary, $S(A|C)$, for the rounds where raw key bits are generated. Using Equation~\eqref{eq:phi_final2_ideal}, the density operator, after dropping off-diagonal terms, with $\ket{k_{i,j}}_{C}\! \langle{k_{l,m}}|,$ for $(i,j) \ne (l,m)$, is
\begin{equation}
\begin{array}{ll}\label{eq:rhof}
\rho_{ABC}=\dfrac{1}{\mathcal{N}} \!\!\! & \Big( \ket{D_v,\! R}_{\!AB}\! \langle{D_v,\! R}| \otimes \ket{k_{0,0}}_{C}\! \langle{k_{0,0}}| + \ket{R,\! D_v}_{\!AB}\! \langle{R,\! D_v}|\otimes \ket{k_{1,1}}_{C}\! \langle{k_{1,1}}| \\[1.5mm]
& \ \  \;\! + \ket{R,\!R}_{\!AB}\! \langle{R,\!R}| \otimes \ket{k_{1,0}}_{C}\! \langle{k_{1,0}}| + \ket{D_v,\!D_v}_{\!AB}\! \langle{D_v,\!D_v}| \otimes \ket{k_{0,1}}_{C}\! \langle{k_{0,1}}| \Big).
\end{array}
\end{equation}
The state $\kb{D_v,\! R}$, describing Alice detecting without a click and Bob reflecting, is associated to a shared key bit $0$. Similarly, $\kb{R,\!D_v}$ is associated to a key bit $1$. Whereas, $\kb{R,\! R}$ and $\kb{D_v,\! D_v}$ corresponds to errors in the key, when the two users establish opposite key bit values.

\red{Now that we have a description of the quantum state, we can use Theorem \ref{thm:cq-entropy}} to compute a bound on the conditional entropy $S(A|C)$ leading us to:
\begin{eqnarray}
\label{eq:cond_entropy_ideal}
S(A|C) \ge \dfrac{\bk{k_{0,0}} + \bk{k_{1,1}}}{\mathcal N}\left[h\left(\frac{\bk{k_{0,0}}}{\bk{k_{0,0}}+\bk{k_{1,1}}}\right) - h(\lambda_0)\right],
\end{eqnarray}
with $\lambda_0$ is defined as in Equation~\eqref{eq:lambda}.

\begin{figure}[H]
  \centering
  \includegraphics[width=0.7\linewidth]{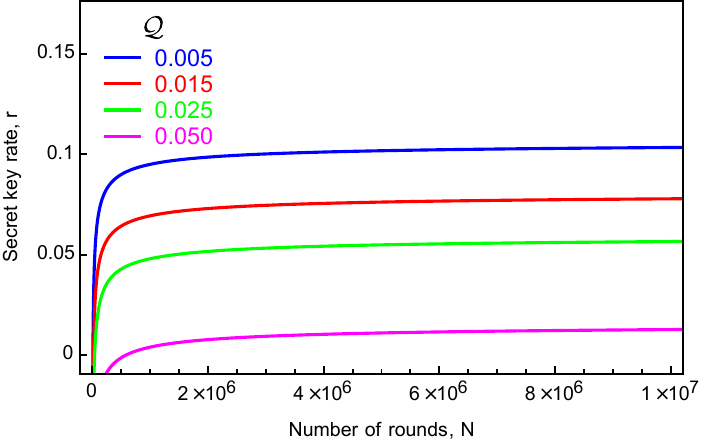}
\caption{\footnotesize The secret key rate $r$ is plotted against $N$, for the ideal case of perfect single-photon sources and detectors. The blue, green and magenta curves correspond to the values of $\mathcal Q$ to be $0.005,0.025$ and $0.05$, respectively. Whereas, the red curve represents the experimentally observed value of $\mathcal Q$, $0.015$.}\label{fig:ideal_main}
\end{figure}

\begin{figure}[H]
\centering
\includegraphics[width=0.85\linewidth]{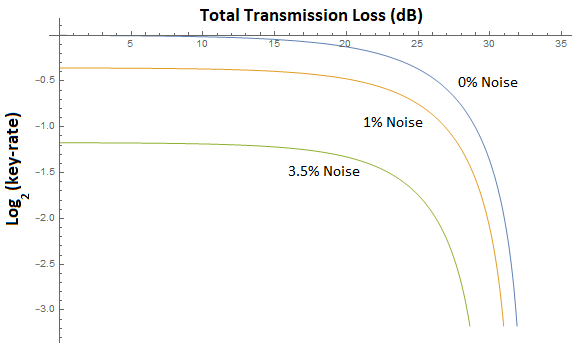}
\caption{\footnotesize The secret key rate $r$ is plotted against transmission loss in one direction (server to end-users), for the ideal case of perfect single-photon sources and detectors in the asymptotic setting (taking the number of iterations to be infinite and with perfect error correction).}\label{fig:ideal_main_loss}
\end{figure}

We present the dependence of the secret key rate $r$ on the total number of rounds $N$ for different values of $\mathcal Q$ (including the one obtained from the experimental set-up) in Figure~\ref{fig:ideal_main} for $T=1$. Other parameters are taken from~\cite{QKD-renner-practical} as $\epsilon = 10^{-5}$, $\epsilon_{EC} = 10^{-10}$ and $\epsilon' = 10^{-7}$. We also assume $\epsilon_{PE} = 10^{-11}$. In Figure \ref{fig:ideal_main_loss}, we report key-rate as a function of total transmission loss in one direction where we set $p_d$ to be a negligible $10^{-8}$ to consider ideal devices on the server also.

\section{Security Analysis - General Case} \label{sec:bound_entropy}
By straightforward algebra, from Equations~\eqref{eq:initial_state}, \eqref{eq:actions_exp} and \eqref{eq:I}, we get $\ket{\phi_2}_{ABSC}=\mathcal I \ket{\phi_1}_{ABfC}$. However, we are only interested in the key-generation rounds, i.e., we condition to the event when the server announces ``1'' and neither Alice nor Bob receives a click. Hence, omitting writing the message state $\ket 1_S$, the final density operator (without the off-diagonal terms) of the system $ABC$ is
\begin{eqnarray}\label{eq:rho_expt}
\!\!\!\!\rho_{ABC}\!=\dfrac{1}{\mathcal{N}} && \Big[ \ket{D_v,\! R}_{\!AB}\! \langle{D_v,\! R}| \otimes \ket{k_{0,0}}_{C}\! \langle{k_{0,0}}| + \ket{R,\! D_v}_{\!AB}\! \langle{R,\! D_v}|\otimes \ket{k_{1,1}}_{C}\! \langle{k_{1,1}}| \nonumber \\
&&\ + \ket{D_\ell,\! R}_{\!AB}\! \langle{D_\ell,\! R}| \otimes \ket{k^1_{0,0}}_{C}\! \langle{k^1_{0,0}}| + \ket{R,\! D_\ell}_{\!AB}\! \langle{R,\! D_\ell}|\otimes \ket{k^1_{1,1}}_{C}\! \langle{k^1_{1,1}}|  \nonumber \\[1mm]
&&\ + \ket{D'_\ell,\! R}_{\!AB}\! \langle{D'_\ell,\! R}| \otimes \ket{k^2_{0,0}}_{C}\! \langle{k^2_{0,0}}| + \ket{R,\! D'_\ell}_{\!AB}\! \langle{R,\! D'_\ell}|\otimes \ket{k^2_{1,1}}_{C}\! \langle{k^2_{1,1}}|  \nonumber \\[1mm]
&&\ + \ket{D_\ell D'_\ell,\! R}_{\!AB}\! \langle{D_\ell D'_\ell,\! R}| \otimes \ket{k^3_{0,0}}_{C}\! \langle{k^3_{0,0}}| + \ket{R,\! D_\ell D'_\ell}_{\!AB}\! \langle{R,\! D_\ell D'_\ell}|\otimes \ket{k^3_{1,1}}_{C}\! \langle{k^3_{1,1}}|   \\[1mm]
&&\ + \ket{D_v,\! D_v}_{\!AB}\! \langle{D_v,\! D_v}| \otimes \ket{k_{0,1}}_{C}\! \langle{k_{0,1}}| + \ket{R,\!R}_{\!AB}\! \langle{R,\!R}| \otimes \ket{k_{1,0}}_{C}\! \langle{k_{1,0}}|  \nonumber\\[1mm]
&&\ + \ket{D_\ell,\! D_v}_{\!AB}\! \langle{D_\ell ,\! D_v}| \otimes \ket{k^1_{0,1}}_{C}\! \langle{k^1_{0,1}}| + \ket{D_v,\! D_\ell}_{\!AB}\! \langle{D_v,\! D_\ell}| \otimes \ket{k^{2}_{0,1}}_{C}\! \langle{k^{2}_{0,1}}|  \nonumber \\[1mm]
&&\ +  \ket{D_\ell,\! D'_\ell}_{\!AB}\! \langle{D_\ell,\! D'_\ell}| \otimes \ket{k^{3}_{0,1}}_{C}\! \langle{k^{3}_{0,1}}| +  \ket{D'_\ell,\! D_\ell}_{\!AB}\! \langle{D'_\ell,\! D_\ell}| \otimes \ket{k^{4}_{0,1}}_{C}\! \langle{k^{4}_{0,1}}|  \nonumber \\
&&\ + \ket{D_\ell D'_\ell,\! D_v}_{\!AB}\! \langle{D_\ell D'_\ell ,\! D_v}| \otimes \ket{k^{5}_{0,1}}_{C}\! \langle{k^{5}_{0,1}}| + \ket{D_v,\! D_\ell D'_\ell}_{\!AB}\! \langle{D_v,\! D_\ell D'_\ell}| \otimes \ket{k^{6}_{0,1}}_{C}\! \langle{k^{6}_{0,1}}| \Big].\nonumber
\end{eqnarray}
Note that, as before, we use commas in the states from $\mathcal H_A \otimes \mathcal H_B$ to separate the quantum numbers defining Alice's and Bob's apparatus states: $\ket{D_\ell D'_\ell,\! R}_{\!AB}$ means that Alice opted to detect, unsuccessfully (due to finite detection efficiency) the two photons present in her lab, while Bob set his apparatus to reflect, etc. The states $\ket{k_{i,j}}_{C}$, etc., are associated to the cases when Alice establishes the value $i$ and Bob $j$ as a key bit, and are given by
\begin{equation}\label{eq:kij}
\begin{array}{ll}
\ket{k_{0,0}} = \dfrac 1 2 \left[ \ket{f_{0,0}^{0,0}} + \ket{f_{0,1}^{0,1}} + \ket{f_{0,2}^{0,2}} \right], \quad\quad\quad\quad & \ket{k_{1,1}} = \dfrac 1 2 \left[\ket{f_{0,0}^{0,0}} + \ket{f_{1,0}^{1,0}} + \ket{f_{2,0}^{2,0}} \right], \\[3mm] 
\ket{k^1_{0,0}} = \dfrac 1 2 \sqrt{\pr_\ell^A} \left[ \ket{f_{0,0}^{1,0}} + \ket{f_{0,1}^{1,1}}\right], \quad &\ket{k^1_{1,1}} = \dfrac 1 2 \sqrt{\pr_\ell^B}  \left[ \ket{f_{0,0}^{0,1}} + \ket{f_{1,0}^{1,1}}\right],  \\[3mm]
\ket{k^2_{0,0}} = \dfrac 1 2 \sqrt{\pr_\ell^A} \ket{f_{0,1}^{1',1}}, \quad & \ket{k^2_{1,1}} = \dfrac 1 2 \sqrt{\pr_\ell^B} \ket{f_{1,0}^{1,1'}},  \\[3mm]
\ket{k^3_{0,0}} = \dfrac 1 2 \pr_\ell^A \ket{f_{0,0}^{2,0}}, \quad & \ket{k^3_{1,1}} = \dfrac 1 2 \pr_\ell^B \ket{f_{0,0}^{0,2}}, \\[6mm]
\ket{k_{0,1}} = \dfrac 1 2 \ket{f_{0,0}^{0,0}}, \quad &\ket{k_{1,0}} = \dfrac 1 2 \big[\ket{f_{0,0}^{0,0}} + \ket{f_{1,0}^{1,0}} + \ket{f_{0,1}^{0,1}} + \ket{f_{2,0}^{2,0}}, \\[3mm]
 \ \ \ \ \ \ \ \ \ \ \ \ \quad  & \ \ \ \ \ \ \ \ \ \ \ \ \ \ \ + \ket{f_{1,1'}^{1,1'}}+ \ket{f_{1',1}^{1',1}}  + \ket{f_{0,2}^{0,2}}\big], \\[6mm]
\ket{k^{1}_{0,1}} = \dfrac 1 2 \sqrt{\pr_\ell^A} \ket{f_{0,0}^{1,0}}, \quad & \ket{k^{4}_{0,1}} = \dfrac 1 2 \sqrt{\pr_\ell^A \;\! \pr_\ell^B} \ket{f_{0,0}^{1',1}}, \\[3mm]
\ket{k^{2}_{0,1}} = \dfrac 1 2 \sqrt{\pr_\ell^B} \ket{f_{0,0}^{0,1}}, \quad & \ket{k^{5}_{0,1}} = \dfrac 1 2 \pr_\ell^A \ket{f_{0,0}^{2,0}}, \\[3mm]
\ket{k^{3}_{0,1}} = \dfrac 1 2 \sqrt{\pr_\ell^A \;\! \pr_\ell^B} \ket{f_{0,0}^{1,1'}}, \quad & \ket{k^{6}_{0,1}} = \dfrac 1 2 \pr_\ell^B \ket{f_{0,0}^{0,2}}.
\end{array}
\end{equation}

Above, as well as in rest of the Appendix, for simplicity we omit writing the labels of the quantum states ($A$, $B$, $C$, $S$ and $f$), whenever it is  implicitly unambiguous to which space they belong by their quantum numbers ($D_v$, $0,0$, etc.).

The normalization constant $\mathcal{N}$ from Equation~\eqref{eq:rho_expt} is the probability to obtain the result ``1'' when there were no clicks at the agents' detectors, given by
\begin{equation}
\begin{array}{ll}\label{eq:p1}
\!\!\mathcal{N} \!=&\!\!\bk{k_{0,0}} \!+\! \bk{k^1_{0,0}} \!+\! \bk{k^2_{0,0}} \!+\! \bk{k^3_{0,0}} \!+\! \bk{k_{1,1}} \!+\! \bk{k^1_{1,1}} \!+\! \bk{k^2_{1,1}} \!+\! \bk{k^3_{1,1}} \\ [2mm]
&\!\!\!\!+\! \bk{k_{0,1}}\!+\! \bk{k^{1}_{0,1}} \!+\! \bk{k^{2}_{0,1}} \!+\! \bk{k^{3}_{0,1}} \!+\! \bk{k^{4}_{0,1}} \!+\! \bk{k^{5}_{0,1}} \!+\! \bk{k^{6}_{0,1}} \!+\! \bk{k_{1,0}}.
\end{array}
\end{equation}
In $\rho_{ABC}$, given by Equation~\eqref{eq:rho_expt}, the state $\kb{D_v,\! R}$ describes Alice detecting without a click and Bob reflecting, and is associated to a shared key bit of $0$. Let us define $\pr_{0,0}=\pr(D_v,\! R \ ; 1)=\bk{k_{0,0}}$ as the joint probability for the event when Alice detects vacuum and Bob reflects, and the server announces the result ``1'', which corresponds to the users sharing a key bit of $0$. Here we use the semicolon (;) to denote logical AND operation between two propositions. Note that $\kb{D_\ell,\! R}$, $\kb{D'_\ell,\! R}$, and $\kb{D_\ell D'_\ell,\! R}$ also correspond to a shared key bit of $0$, and are a consequence of Alice's imperfect detector and multi-photon events. Therefore, one can analogously define the probabilities $\pr^1_{0,0}, \pr^2_{0,0}$ and $\pr^3_{0,0}$, such that the total probability of the users sharing a key bit of $0$ can be given by $\tilde{\pr}_{0,0}=\pr_{0,0} + \pr^1_{0,0} + \pr^2_{0,0} + \pr^3_{0,0}$. Analogously, the probabilities, $\pr_{1,1}, \pr^1_{1,1}, \pr^2_{1,1}$ and $\pr^3_{1,1}$, associated to a key bit $1$ are defined. The $k_{ij}$'s with $i\neq j$ are associated to the errors, i.e., when the two users establish opposite key bit values. From the above definitions, using $k_{ij}$ and $\mathcal N$, we have
\begin{equation}
\dfrac{\bk{k_{0,0}}+ \bk{k^1_{0,0}} + \bk{k^2_{0,0}} + \bk{k^3_{0,0}}}{\mathcal N}=\pr(D_v,\! R \vee D_\ell,\! R \vee D'_\ell,\! R \vee D_\ell D'_\ell,\! R| 1).
\end{equation}
Here, by $\pr(\mathcal P|\mathcal C)$ we denote the conditional probability that the proposition $\mathcal P$ holds (in the above case, Alice detects and observes no clicks, while Bob reflects), given that the condition $\mathcal C$ is satisfied (in the above case, the server announces ``1''). Therefore, using the following terminology for different probabilities (to be used in parameter estimation described in the next section), the probability to share the key is given by
\begin{eqnarray}\label{eq:pkey}
\pr_{key} \!&=& \left[\bk{k_{0,0}} \!+\! \bk{k^1_{0,0}} \!+\! \bk{k^2_{0,0}} \!+\! \bk{k^3_{0,0}}\right] \!+\! \left[\bk{k_{1,1}} \!+\! \bk{k^1_{1,1}} \!+\! \bk{k^2_{1,1}} \!+\! \bk{k^3_{1,1}}\right] \nonumber\\ [1mm]
&=&\left[\pr_{0,0} + \pr^1_{0,0} + \pr^2_{0,0} + \pr^3_{0,0}\right] + \left[\pr_{1,1} + \pr^1_{1,1} + \pr^2_{1,1} + \pr^3_{1,1}\right] \nonumber\\ [1mm]
&=& \tilde{\pr}_{0,0} + \tilde{\pr}_{1,1} \nonumber\\ [1mm]
&=& \pr(D_v,\! R \vee D_\ell,\! R  \vee D'_\ell,\! R \vee D_\ell D'_\ell,\! R \ \! ; 1) + \pr(R,\! D_v \vee R,\!D_\ell \vee R,\!D'_\ell \vee R,\! D_\ell D'_\ell \ \! ; 1),
\end{eqnarray}
where $\pr(D_v,\! R \vee D_\ell,\! R \vee D'_\ell,\! R \vee D_\ell D'_\ell,\! R  \  ; 1)$ represents the joint probability of the following event: Alice detects vacuum, Bob reflects, and the server announces the result ``1''; and analogously for the other term. As before, we use the semicolon (;) to denote logical AND operation between two propositions, instead of introducing the additional parenthesis for the first one, and using the standard symbol $\wedge$. The probability of error in the raw key is given by
\begin{eqnarray}\label{eq:perror}
\pr_{err} \!\!&=&\left[\bk{k_{0,1}} \!+\! \bk{k^{1}_{0,1}} \!+\! \bk{k^2_{0,1}} \!+\!  \bk{k^3_{0,1}} \!+\! \bk{k^{4}_{0,1}} \!+\! \bk{k^{5}_{0,1}} \!+\! \bk{k^{6}_{0,1}}\right] \!+\! \bk{k_{1,0}} \nonumber\\ [1mm]
&=&\left[ \pr_{0,1} + \pr^{1}_{0,1} + \pr^2_{0,1} + \pr^3_{0,1} + \pr^{4}_{0,1} + \pr^{5}_{0,1} + \pr^{6}_{0,1}\right] + \pr_{1,0}\nonumber\\ [1mm]
&=&\tilde{\pr}_{0,1} + \tilde{\pr}_{1,0} \nonumber\\ [1mm]
&=&\pr(D_v,\!D_v \vee D_\ell ,\!D_v \vee D_v,\! D_\ell \vee D_\ell,\! D'_\ell \vee D'_\ell,\! D_\ell \vee D_v,\! D_\ell D'_\ell \vee D'_\ell D_\ell,\! D_v  \ \! ; 1) + \pr(R R  \ \! ; 1),
\end{eqnarray}
where $\pr(D_v,\!D_v \vee D_\ell ,\!D_v \vee D_v,\! D_\ell \vee D_\ell,\! D'_\ell \vee D'_\ell,\! D_\ell \vee D_v,\! D_\ell D'_\ell \vee D'_\ell D_\ell,\! D_v  \ \! ; 1)$ represents the joint probability of the event: Alice and Bob both detect vacuum, and that the server announces the result ``1''; and analogously for the other term. Note that the probabilities $\tilde{\pr}_{i,j}$ can be observed from the experiment directly.

To obtain the secret key rate, we again use the bound given in Theorem \ref{thm:cq-entropy}, as
\begin{equation}
\begin{array}{ll}\label{eq:cond_entropy}
S(A|C) &\ge \dfrac{\bk{k_{0,0}} + \bk{k_{1,1}}}{\mathcal N}\left(h\left[\dfrac{\bk{k_{0,0}}}{\bk{k_{0,0}}+\bk{k_{1,1}}}\right] - h(\lambda_0)\right)\\[5mm]
&+\; \dfrac{\bk{k^1_{0,0}} + \bk{k^1_{1,1}}}{\mathcal N}\left(h\left[\dfrac{\bk{k^1_{0,0}}}{\bk{k^1_{0,0}}+\bk{k^1_{1,1}}}\right] - h(\lambda_1)\right)\\[5mm]
&+\; \dfrac{\bk{k^2_{0,0}} + \bk{k^2_{1,1}}}{\mathcal N}\left(h\left[\dfrac{\bk{k^2_{0,0}}}{\bk{k^2_{0,0}}+\bk{k^2_{1,1}}}\right] - h(\lambda_2)\right)\\[5mm]
&+\; \dfrac{\bk{k^3_{0,0}} + \bk{k^3_{1,1}}}{\mathcal N}\left(h\left[\dfrac{\bk{k^3_{0,0}}}{\bk{k^3_{0,0}}+\bk{k^3_{1,1}}}\right] - h(\lambda_3)\right)\\[5mm]
&+\; \dfrac{\bk{k_{0,1}} + \bk{k_{1,0}}}{\mathcal N}\left(h\left[\dfrac{\bk{k_{0,1}}}{\bk{k_{0,1}}+\bk{k_{1,0}}}\right] - h(\lambda_4)\right),
\end{array}
\end{equation}
where $h(\cdot )$ is the binary Shannon entropy, and $\lambda_{i}$'s are defined in the following way
\begin{equation}
\begin{array}{ll}\label{eq:lambda}
\lambda_0 &= \dfrac{1}{2}\left(1 + \dfrac{\sqrt{\left(\bk{k_{0,0}}-\bk{k_{1,1}}\right)^2 + 4\re^2\braket{k_{0,0}|k_{1,1}}}}{\bk{k_{0,0}}+\bk{k_{1,1}}}\right),\\[6mm]
\lambda_1 &= \dfrac{1}{2}\left(1 + \dfrac{\sqrt{\left(\bk{k^1_{0,0}}-\bk{k^1_{1,1}}\right)^2 + 4\re^2\braket{k^1_{0,0}|k^1_{1,1}}}}{\bk{k^1_{0,0}}+\bk{k^1_{1,1}}}\right),\\[6mm]
\lambda_2 &= \dfrac{1}{2}\left(1 + \dfrac{\sqrt{\left(\bk{k^2_{0,0}}-\bk{k^2_{1,1}}\right)^2 + 4\re^2\braket{k^2_{0,0}|k^2_{1,1}}}}{\bk{k^2_{0,0}}+\bk{k^2_{1,1}}}\right),\\[6mm]
\lambda_3 &= \dfrac{1}{2}\left(1 + \dfrac{\sqrt{\left(\bk{k^3_{0,0}}-\bk{k^3_{1,1}}\right)^2 + 4\re^2\braket{k^3_{0,0}|k^3_{1,1}}}}{\bk{k^3_{0,0}}+\bk{k^3_{1,1}}}\right),\\[6mm]
\lambda_4 &= \dfrac{1}{2}\left(1 + \dfrac{\sqrt{\left(\bk{k_{0,1}}-\bk{k_{1,0}}\right)^2 + 4\re^2\braket{k_{0,1}|k_{1,0}}}}{\bk{k_{0,1}}+\bk{k_{1,0}}}\right).
\end{array}
\end{equation}
The first four terms in $S(A|C)$ correspond to the keys shared between Alice and Bob, while the last term corresponds to errors in the key. However, we estimate the lower bound on $S(A|C)$ by considering only the first term since its contribution to the entropy is far larger than that of any of the other terms. From the expression~\eqref{eq:lambda} for $\lambda_0$, we see that minimizing $S(A|C)$ essentially means minimizing $\re \braket{k_{0,0}|k_{1,1}}$. Therefore, in addition to different probabilities obtained from the experiment, we also need to estimate $\re \braket{k_{0,0}|k_{1,1}}$. We proceed by computing the lower bound for $\re^2 \braket{k_{0,0}|k_{1,1}}$, i.e., for $|\re \braket{k_{0,0}|k_{1,1}}|$. Notice that the lower it is, the closer to 1/2 $\lambda_0$ is, i.e., the closer to 1 the $h(\lambda_0)$ is, and the worst case scenario for $S(A|C)$, has the lowest value.

Let us use the following notation for simplification,
\begin{equation}
\ket x = \ket{f_{1,0}^{1,0}} + \ket{f_{2,0}^{2,0}}, \quad \quad \ \ \ket y = \ket{f_{0,1}^{0,1}} + \ket{f_{0,2}^{0,2}}, \quad \quad \ \ \ket z = \ket{f_{1,1'}^{1,1'}}+ \ket{f_{1',1}^{1',1}}.
\end{equation}
We can rewrite $\ket{k_{0,0}}$ and $\ket{k_{1,1}}$ from Equation~\eqref{eq:kij}, to obtain $\re \braket{k_{0,0}|k_{1,1}}$ as
\begin{equation}\label{eq:overlap}
\re \braket{k_{0,0}|k_{1,1}} = \dfrac 1 4 \left[\bk{f_{0,0}^{0,0}} +\re \braket{x|f_{0,0}^{0,0}} + \re\braket{f_{0,0}^{0,0}|y} + \re\braket{x|y}\right].
\end{equation}
From the error term, we have $\bk{k_{1,0}}=\pr(R,\!R|1) \pr(R,\!R)=\mathcal Q /4$, where $\mathcal Q = \pr(R,\!R|1)$ is the probability that the server announces the result ``1'', given both Alice and Bob reflected, and $\pr(R,\!R)=1/4$. With straightforward substitution from the above into Equation~\eqref{eq:overlap}, with $\bk{f_{0,0}^{0,0}}=4\bk{k_{0,1}}=4\pr_{0,1}$, we get
\begin{equation}
\braket{k_{0,0}|k_{1,1}} = \dfrac{\mathcal Q}{8} + \dfrac{\pr_{0,1}}{2} - \dfrac{1}{8}\left[\bk{x}+\bk{y}+\bk{z}\right] - \dfrac 1 4 \left[ \braket{x|z} + \braket{y|z} + \braket{f_{0,0}^{0,0}|z} \right].
\end{equation}
In the ideal case, with no vacuum or multi-photon pulses, when $\bk{x} = 4 \bk{k_{0,0}}= 4 \pr_{0,0}$ and $\bk{y} = 4 \bk{k_{1,1}}= 4 \pr_{1,1}$, we recover the ideal case expression~\ref{eq:ideal_overlap} from Appendix~\ref{sec:ideal}. By writing $\braket{x|z} = |\braket{x|z}| e^{\varphi_{x,z}}$, we have 
\begin{equation}
\re\braket{x|z} = |\braket{x|z}| \cos\varphi_{x,y} = ||\ket{x}|| \cdot ||\ket{z}|| \cdot |\cos{\chi_{x,z}}|\cos\varphi_{x,z} = \sqrt{\bk{x}} \sqrt{\bk{z}} \cos{\theta_{x,z}}, 
\end{equation}
where $\chi_{x,z}$ denotes the angle between $\ket x$ and $\ket z$ and $\cos{\theta_{x,z}} \equiv |\cos{\chi_{x,z}}|\cos\varphi_{x,z}$, and analogously for $\re\braket{y|z}$ and so on. Therefore, the final expression for $\re \braket{k_{0,0}|k_{1,1}}$ is
\begin{eqnarray}\label{eq:overlap_final}
\re \braket{k_{0,0}|k_{1,1}}&=& \dfrac{\mathcal Q}{8} + \dfrac{\pr_{0,1}}{2} - \dfrac{1}{8}\left[\bk{x}+\bk{y}+\bk{z}\right] - \dfrac 1 4 \left[\sqrt{\bk{f_{0,0}^{0,0}}} \sqrt{\bk{z}} \cos{\theta_{f,z}} \right]\nonumber\\[1mm]
&& - \dfrac 1 4 \left[ \sqrt{\bk{x}} \sqrt{\bk{z}} \cos{\theta_{x,z}} + \sqrt{\bk{y}} \sqrt{\bk{z}} \cos{\theta_{y,z}} \right].
\end{eqnarray}

To obtain $\bk x$ and $\bk y$, consider again $\ket{k_{0,0}}$ and $\ket{k_{1,1}}$ from Equation~\eqref{eq:kij}
\begin{equation}\label{eq:xx_yy}
\begin{array}{ll}
\bk{k_{1,1}}\!= \! \dfrac 1 4 \left[\bk{f_{0,0}^{0,0}} + \bk x + 2 \re\braket{f_{0,0}^{0,0}|x} \right], \\[3mm] \bk{k_{0,0}}\!= \! \dfrac 1 4 \left[\bk{f_{0,0}^{0,0}} + \bk y + 2 \re\braket{f_{0,0}^{0,0}|y} \right].
\end{array}
\end{equation}
Note that $\bk{f_{0,0}^{0,0}}=4\bk{k_{0,1}}=4\pr_{0,1}, \bk{k_{0,0}}=\pr_{0,0}$ and $\bk{k_{1,1}}=\pr_{1,1}$. Therefore, solving the quadratic equations obtained from~\eqref{eq:xx_yy}, we get the following positive roots of $\sqrt{\bk x}$ and $\sqrt{\bk y}$,
\begin{equation}
\begin{array}{ll}
\sqrt{\bk x} &= 2 \left[- \sqrt{\pr_{0,1}} \; \mbox{cos}\;\! \theta_{x,f} + \sqrt{\pr_{1,1}-(1-\; \mbox{cos}^2 \theta_{x,f})\;\pr_{0,1}}\;\right],\\[2mm]
\sqrt{\bk y} &= 2 \left[- \sqrt{\pr_{0,1}} \; \mbox{cos}\;\! \theta_{y,f} + \sqrt{\pr_{0,0}-(1-\; \mbox{cos}^2 \theta_{y,f})\;\pr_{0,1}}\;\right].
\end{array}
\end{equation}
Analogously, for $\bk z$ we have
\begin{equation}
\begin{array}{ll}
&\!\!\!\!\!\!\!\!\!\!\!\! \bk z + \underbrace{2 \left[\sqrt{\bk x} \mbox{ cos } \theta_{x,z} + \sqrt{\bk y} \mbox{ cos } \theta_{y,z} + 2\sqrt{\pr_{0,1}} \mbox{ cos } \theta_{f,z} \right]}_{\beta} \sqrt{\bk z} \\
& \ \ \ \ \ \ \ + \ 4\left[\pr_{0,1} -\pr_{1,0}\right] + \left[\bk x + \bk y + 2 \sqrt{\bk x} \sqrt{\bk y} \mbox{ cos } \theta_{x,y}\right] \\[1mm]
& \ \ \ \ \ \ \ \ \ \ \underbrace{ \ \ \ \ \ \ \ \ \ \ \ \ \ \ \  \ \ \ \ \ \ \ \ \ \ \ \ \ \ \ \ \ \ \ + \ 4 \sqrt{\pr_{0,1}} \left[ \sqrt{\bk x} \mbox{ cos } \theta_{x,f}+\sqrt{\bk y} \mbox{ cos } \theta_{y,f}\right]}_{\gamma}\!=0,
\end{array}
\end{equation}
where $\cos{\theta_{x,z}} \equiv |\cos{\chi_{x,z}}|\cos\varphi_{x,z}$ and analogously for $\cos{\theta_{y,z}}$, $\cos{\theta_{f,z}}$, etc.
Again, solving the above quadratic equation, we can obtain the positive root of $\sqrt{\bk z}$.

\section{Parameter estimation}
\label{subsec:parameter_estimation}

Here, we briefly explain how to estimate the relevant probabilities, $\pr_{0,0},\pr_{1,1}$ and $\pr_{0,1}$, to compute $S(A|C)$ in Equation~\eqref{eq:cond_entropy}, to eventually obtain the secret key rate given by Equation~(1) from the main text.

Due to the nature of this protocol, in the ideal case, one expects $\pr (1) = 1/8$ (see Appendix~\ref{sec:ideal} for details), which is further reduced in the experimental case of imperfect detectors, etc. Therefore, it is useful if these probabilities could be computed without sacrificing any key-generation rounds. Below, we discuss the case with direct estimation where Alice and Bob use part of the key to obtain these probabilities, as well as the case of indirect estimation where no key-generation rounds are wasted.

\subsection{Direct estimation}\label{subsubsec:direct_estimation}

Here, we sacrifice $\mu$ instances of the total $N_{raw}$ key-generation rounds, to directly compute the relevant probabilities. However, since Alice's and Bob's detectors are imperfect, they cannot compute $\pr_{0,0}=\pr(D_v,\! R \ \! ; 1)$ and $\pr_{1,1}=\pr(R,\!D_v \ \! ; 1)$ directly, as they cannot differentiate the event $D_v,\!R$ from the events $D_\ell,\!R$, $D'_\ell,\!R$ and $D_\ell D'_\ell,\!R$, and analogously for $R,\!D_v$. However, they can obtain $\tilde{\pr}_{0,0}=\pr_{0,0}+\pr^1_{0,0}+\pr^2_{0,0} +\pr^3_{0,0}=\pr(D_v,\! R \vee D_\ell,\!R \vee D'_\ell,\!R \vee D_\ell D'_\ell,\!R \ \! ; 1)$ directly, and also $\tilde{\pr}_{1,1}$. They can then compute $\pr^1_{0,0}=\bk{k^1_{0,0}}$, $\pr^2_{0,0}=\bk{k^2_{0,0}}$ and $\pr^3_{0,0}=\bk{k^3_{0,0}}$, to eventually obtain $\pr_{0,0}$. From Equation~\eqref{eq:kij} one has
\begin{eqnarray}
&&\pr^1_{0,0}=\pr(D_\ell,\!R \ \! ; 1)=\dfrac{\pr_\ell^A}{4} \left( ||\ket{f_{0,0}^{1,0}} + \ket{f_{0,1'}^{1,1'}}||^2 \right), \nonumber \\
&& \pr^2_{0,0}=\pr(D'_\ell,\!R \ \! ; 1)=\dfrac{\pr_\ell^A}{4} \bk{f_{0,1}^{1',1}},\nonumber \\
&&\pr^3_{0,0}=\pr(D_\ell D'_\ell,\!R \ \! ; 1)=\dfrac{\pr_\ell^{A^2}}{4} \bk{f_{0,0}^{2,0}}.
\end{eqnarray}
Even though Alice and Bob cannot compute the above probabilities, they can estimate them by looking at the events corresponding to the clicks, using the expressions
\begin{eqnarray}\label{eq:D_cR}
&&\pr(D_c,\!R \ \! ; 1)=\dfrac{\pr_d^A}{4} \left( ||\ket{f_{0,0}^{1,0}} + \ket{f_{0,1'}^{1,1'}}||^2 \right), \nonumber \\
&& \pr(D'_c,\!R \ \! ; 1)=\dfrac{\pr_d^A}{4} \bk{f_{0,1}^{1',1}}, \nonumber \\
&&\pr(D_c D'_c,\!R \ \! ; 1)=\dfrac{\pr_d^{A^2}}{4} \bk{f_{0,0}^{2,0}}.
\end{eqnarray}
Therefore, we can write $\left(\pr^1_{0,0} + \pr^2_{0,0}\right)$ and $\pr^3_{0,0}$ as
\begin{align}
&\pr^1_{0,0} + \pr^2_{0,0} = \left(\dfrac{\pr_\ell^A}{\pr_d^A}\right) \pr(D_c,\!R \vee D'_c,\!R \ \! ; 1), &&\pr^3_{0,0} = \left(\dfrac{\pr_\ell^{A}}{\pr_d^{A}}\right)^{\!\!\!2} \pr(D_c D'_c,\!R \ \! ; 1),\label{eq:47}
\end{align}
where only $\pr(D_c D'_c,\!R \ \! ; 1)$ can be obtained using the rounds when Alice gets double clicks in her detector. However, $\pr(D_c,\!R \vee D'_c,\!R \vee D_\ell D'_c,\!R \vee D_c D'_\ell,\!R \ \! ; 1)$, corresponding to a single click in Alice's detector, can also be obtained directly. Hence,
\begin{equation}
\pr(D_c,\!R \vee D'_c,\!R \ \! ; 1)=\pr(D_c,\!R \vee D'_c,\!R \vee D_\ell D'_c,\!R \vee D_c D'_\ell,\!R \ \! ; 1) - \pr(D_\ell D'_c,\!R \ \! ; 1) - \pr(D_c D'_\ell,\!R \ \! ; 1).
\end{equation}
Also, we have
\begin{equation}
\pr(D_\ell D'_c,\!R \ \! ; 1)= \dfrac{\pr_\ell^A \pr_d^A}{4} \bk{f_{0,0}^{2,0}} = \pr(D_c D'_\ell,\!R \ \! ; 1).
\end{equation}
Therefore, the required probabilities $\pr_{0,0}$ and $\pr_{1,1}$ are
\begin{equation}
\begin{array}{ll}\label{eq:p00p11_direct}
&\pr_{0,0} = \tilde{\pr}_{0,0} - \left(\dfrac{\pr_\ell^A}{\pr_d^A}\right) \pr(D_c,\!R \vee D'_c,\!R \vee D_\ell D'_c,\!R \vee D_c D'_\ell,\!R \ \! ; 1) + \left(\dfrac{\pr_\ell^{A}}{\pr_d^{A}}\right)^{\!\!\!2} \pr(D_c D'_c,\!R \ \! ; 1), \\[3.5mm]
&\pr_{1,1} = \tilde{\pr}_{1,1} - \left(\dfrac{\pr_\ell^B}{\pr_d^B}\right) \pr(R,\!D_c \vee R,\!D'_c \vee R,\! D_\ell D'_c \vee R,\! D_c D'_\ell \ \! ; 1) + \left(\dfrac{\pr_\ell^{B}}{\pr_d^{B}}\right)^{\!\!\!2} \pr(D_c D'_c,\!R \ \! ; 1).
\end{array}
\end{equation}
Additionally, to compute $\pr_{0,1}$, required to estimate $\re \braket{k_{0,0}|k_{1,1}}$ from Equation~\eqref{eq:overlap_final}, we use $\pr_{0,1}=\tilde{\pr}_{0,1} - \pr^{1}_{0,1} - \pr^2_{0,1} - \pr^3_{0,1}  - \pr^{4}_{0,1} - \pr^{5}_{0,1} - \pr^{6}_{0,1}$. Again, using straightforward algebra, we have
\begin{eqnarray}\label{eq:p01_direct}
\pr_{0,1} &=& \tilde{\pr}_{0,1} - \left(\dfrac{\pr_\ell^A}{\pr_d^A}\right) \pr(D_c,\!D_v \vee D_c,\!D'_\ell \vee D'_c,\!D_\ell \vee D_c D'_\ell,\!D_v \vee D_\ell D'_c,\! D_v \ \! ;  1) \nonumber \\
&& \ \ \ \ - \left(\dfrac{\pr_\ell^B}{\pr_d^B}\right) \pr(D_v,\!D_c \vee D_\ell,\!D'_c \vee D'_\ell,\!D_c \vee D_v,\!D_c D'_\ell \vee D_v,\!D_\ell D'_c \ \! ; 1)  \\
&& \ \ \ \ -\; 3 \left(\dfrac{\pr_\ell^{A}}{\pr_d^{A}}\right)^{\!\!\!2} \pr(D_c D'_c,\! D_v \ \! ; 1)- 3 \left(\dfrac{\pr_\ell^{B}}{\pr_d^{B}}\right)^{\!\!\!2} \pr(D_v,\!D_c D'_c \ \! ; 1) - 3 \left(\dfrac{\pr_\ell^A \;\! \pr_\ell^B}{\pr_d^A \;\! \pr_d^B}\right) \pr(D_c,\!D'_c \vee D'_c,\!D_c \ \! ; 1).\nonumber
\end{eqnarray}

\begin{figure}
  \centering
  \includegraphics[width=0.7\linewidth]{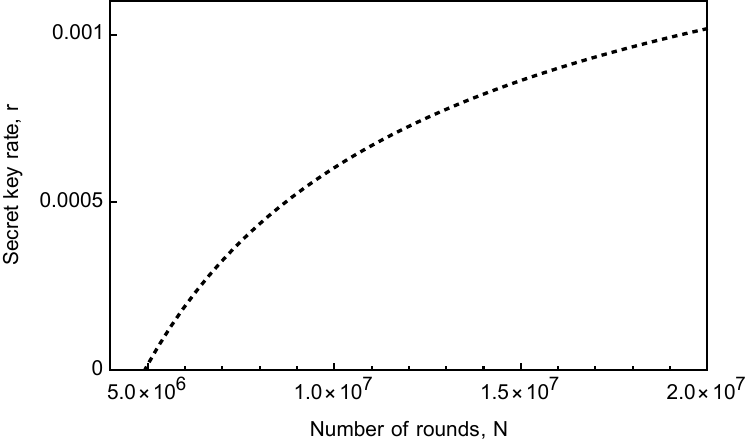}
\caption{Secret key rate, $r$, vs number of rounds, $N$, for the case of imperfect single-photon sources and detectors. The probability necessary for the plot are obtained from the experimental data.}\label{fig:fixed_efficiency}
\end{figure}

Using the direct estimation method to compute all the relevant probabilities, we obtain the secret key rate $r$ (from Equation~(1) from the main text) in Figure~\ref{fig:fixed_efficiency}. We consider the implemented number of rounds, $10^5$, as a subset of a larger implementation and, therefore, use them to estimate the secret key rate. The probability of server announcing ``1'' during these rounds is $\pr(1)=0.0162$. Therefore, the amount of keys wasted during the parameter estimations is $1620$ bits.

The probabilities of Equations~\eqref{eq:p00p11_direct} and \eqref{eq:p01_direct} are the following: $\pr_{0,0}=(7.3 \pm 0.3) \times 10^{-3}$, $\pr_{1,1} = (5.5 \pm 0.3) \times 10^{-3}$, $\pr_{0,1}=(1.1 \pm 0.9 ) \times 10^{-4}$ and $\pr_{1,0} = (5.1 \pm 0.7) \times 10^{-4}$. 

We assume $\epsilon = 10^{-5}$, $\epsilon_{EC} = 10^{-10}$ and $\epsilon_{PE} = 10^{-11}$. 
The value $\epsilon'$ is a factor in the min-entropy expression used for the key rate computation and may actually be set by the user arbitrarily to maximize the key rate (see Lemma 1 from~\cite{QKD-renner-practical}). However, for our evaluations we simply set $\epsilon^\prime = 10^{-7}$ (optimizing this could only improve our results). For parameter estimation, we take $\epsilon_{PE}=10^{-11}$ and assume a confidence interval $\delta=10^{-4}$, given our experimental errors. The calculated secret key rate corresponds to the minimum lower bound of the entropy $S(A|C)$ (see Equation \eqref{eq:cond_entropy}) over the confidence interval of the experimental probabilities. This minimum occurs for the highest value of the error probability $\pr_{err}$ and the lowest of $\pr_{key}$, and therefore represents the worst possible key rate within our experimental uncertainty.


\subsection{Indirect estimation}\label{subsubsec:indirect_estimation}

To avoid wasting the rounds used for key-generation (when ``1'' was announced without any clicks at Alice's and Bob's detectors), we can use the remaining rounds (when ``0'',``$v$'' or ``$m$'' was announced or ``1'' was announced with click(s) at Alice's and Bob's detectors) for parameter estimation. For these cases, Alice and Bob can communicate over an authenticated channel to convey their respective action choices and resulting states to each other. Therefore, they can communicate for the non-useful rounds where server announces ``0'',``$v$'' or ``$m$'', as well as the rounds where any of them detects a photon in case the server announces ``1''. This method can be applied also in the ideal case described in Section~\ref{sec:ideal}, but we present it only once for brevity.

We know that $\pr_{0,0}=\pr(D_v,\! R \ \! ;  1) = \pr(D_v,\!R) - \pr(D_v,\!R \ \! ; 0) - \pr(D_v,\!R \ \! ; v) - \pr(D_v,\!R \ \! ; m)$, where $\pr(D_v,\! R) = \pr(D\!,\!R) - \pr(D_\ell,\!R) - \pr(D'_\ell,\!R) - \pr(D_c,\! R) - \pr(D'_c,\! R) - \pr(D_\ell D'_\ell,\!R) - \pr(D_c D'_\ell,\!R) - \pr(D_\ell D'_c,\!R) - \pr(D_c D'_c,\!R)$. Note that $\pr(D\!,\!R)$ is the probability of Alice choosing to detect and Bob to reflect. Since Alice and Bob choose their actions at random, ideally $\pr(D\!,\!D)=\pr(D\!,\!R)=\pr(R,\!D)=\pr(R,\!R)=1/4$. However, considering the finite sample size and the inefficiency of switching between the actions, Alice and Bob do not take these probabilities to be $1/4$ but compute them considering only the non-useful rounds. Therefore, we have
\begin{eqnarray}
\pr_{0,0} &=& \pr(D\!,\!R) - \pr(D_\ell,\!R) - \pr(D'_\ell,\!R) - \pr(D_c,\! R) - \pr(D'_c,\! R) - \pr(D_\ell D'_\ell,\!R) - \pr(D_c D'_\ell,\!R) \nonumber\\
&& -  \; \pr(D_\ell D'_c,\!R) - \pr(D_c D'_c,\!R) - \pr(D_v,\!R \ \! ; 0) - \pr(D_v,\!R \ \! ; v) - \pr(D_v,\!R \ \! ; m).
\end{eqnarray}
Note that Alice and Bob cannot directly compute all the quantities from the above expression, say, $\pr(D_v,\!R \ \! ; 0)$, $\pr(D_v,\!R \ \! ; 1)$, etc. They can compute $\pr(D_\ell,\!R), \pr(D'_\ell,\!R)$ and $\pr(D_c D'_\ell,\!R)$ analogously as in the previous subsection, given by Equation~\eqref{eq:47}. However, $\pr(D_c,\!R \vee D'_c,\!R \vee D_\ell D'_c,\!R \vee D_c D'_\ell,\!R)$ can be computed directly. We use $\pr(D_v,\!R \vee D_\ell,\!R \vee D'_\ell,\!R \vee D_\ell D'_\ell,\!R \ \! ; 0)$, directly observable, to estimate $\pr(D_v,\!R \ \! ; 0)$. Therefore,
\begin{eqnarray}\label{eq:MvR}
\!\!\!\!\!\!\!\!\!\!\!\!\!\!\!\!\!\!\pr(D_v,\!R\ \! ; 0)&=&\pr(D_v,\!R \vee\! D_\ell,\!R \vee\! D'_\ell,\!R \vee\! D_\ell D'_\ell,\!R \ \! ; 0)\!-\pr(D_\ell,\!R \ \! ; 0)\! -\pr(D'_\ell,\!R \ \! ; 0)\! -\pr(D_\ell D'_\ell,\!R \ \! ; 0)\!,
\end{eqnarray}
$\pr(D_\ell,\!R \ \! ; 0), \pr(D'_\ell,\!R \ \! ; 0)$ and $\pr(D_c D'_\ell,\!R \ \! ; 0)$, etc., can again be computed in the same way as before. Therefore, the final expressions for $\pr_{0,0}$ and $\pr_{1,1}$, in terms of probabilities computed indirectly, are
\begin{eqnarray}
\pr_{0,0}&&=\pr(D\!,\!R)\! - \pr(D_c,\!R \vee \!D'_c,\!R \vee \!D_\ell D'_c,\!R \vee \!D_c D'_\ell,\!R)\! - \pr(D_c D'_c,\! R)\! -\pr(D_v,\!R \vee\! D_\ell,\!R \vee\! D'_\ell,\!R \vee\! D_\ell D'_\ell,\!R \ \! ; 0)\nonumber\\
&&\ \ \ - \; \pr(D_v,\!R \vee D_\ell,\!R \vee D'_\ell,\!R \vee D_\ell D'_\ell,\!R \ \! ; v) -  \pr(D_v,\!R \vee D_\ell,\!R \vee D'_\ell,\!R \vee D_\ell D'_\ell,\!R \ \! ; m)\nonumber \\
&&\ \ \ + \left(\dfrac{\pr_\ell^A}{\pr_d^A}\right) \big[\pr(D_c,\!R \vee D'_c,\!R \vee D_\ell D'_c,\!R \vee D_c D'_\ell,\!R \ \! ; 0) + \pr(D_c,\!R \vee D'_c,\!R \vee D_\ell D'_c,\!R \vee D_c D'_\ell,\!R \ \! ; v)\nonumber \\
&&\ \ \ + \;\pr(D_c,\!R \vee D'_c,\!R \vee D_\ell D'_c,\!R \vee D_c D'_\ell,\!R \ \! ; m) -\pr(D_c,\!R \vee D'_c,\!R \vee D_\ell D'_c,\!R \vee D_c D'_\ell,\!R) \big]\nonumber \\
&&\ \ \ - \left(\dfrac{\pr_\ell^A}{\pr_d^A}\right)^{\!\!2} \left[\pr(D_c D'_c,\!R \ \! ; 0) + \pr(D_cD'_c,\!R \ \! ; v) +\pr(D_cD'_c,\!R \ \! ; m) -\pr(D_cD'_c,\! R) \right],
\end{eqnarray}
\begin{eqnarray}
\pr_{1,1}&&=\pr(R,\!D)\! - \pr(R,\!D_c \! \vee\! R,\!D'_c \vee\! R,\!D_\ell D'_c \vee\! R,\!D_c D'_\ell)\! - \pr(R,\!D_c D'_c)\! -\pr(R,\!D_v \vee\! R,\!D_\ell \vee \!R,\!D'_\ell \vee\! R,\!D_\ell D'_\ell \ \! ; 0)\nonumber \\
&&\ \ \ - \; \pr(R,\!D_v \vee R,\!D_\ell \vee R,\!D'_\ell \vee R,\!D_\ell D'_\ell \ \! ; v) -  \pr(R,\!D_v \vee R,\!D_\ell \vee R,\!D'_\ell \vee R,\!D_\ell D'_\ell \ \! ; m)\nonumber \\
&& \ \ \ +  \left(\dfrac{\pr_\ell^B}{\pr_d^B}\right) \big[\pr(R,\!D_c \vee R,\!D'_c \vee R,\!D_\ell D'_c \vee R,\!D_c D'_\ell \ \! ; 0) + \pr(R,\!D_c \vee R,\!D'_c \vee R,\!D_\ell D'_c \vee R,\!D_c D'_\ell \ \! ; v)\nonumber \\
&&\ \ \ + \; \pr(R,\!D_c \vee R,\!D'_c \vee R,\!D_\ell D'_c \vee R,\!D_c D'_\ell \ \! ; m) -\pr(R,\!D_c \vee R,\!D'_c \vee R,\!D_\ell D'_c \vee R,\!D_c D'_\ell) \big]\nonumber \\
&&\ \ \ - \left(\dfrac{\pr_\ell^B}{\pr_d^B}\right)^{\!\!2} \left[\pr(R,\!D_c D'_c \ \! ; 0) + \pr(R,\!D_cD'_c \ \! ; v) +\pr(R,\!D_cD'_c \ \! ; m) -\pr(R,\!D_cD'_c) \right].
\end{eqnarray}

We can analogously estimate $\pr_{0,1}$ by computing $\tilde{\pr}_{1,0}$ as
\begin{equation}
\tilde{\pr}_{1,0}=\pr(R,\!R \ \! ; 1) = \pr(RR) - \pr(R,\!R \ \! ; 0) - \pr(R,\!R \ \! ; v) - \pr(R,\!R \ \! ; m).
\end{equation}
Therefore,
\begin{eqnarray}
\pr_{0,1}&&=\pr(1) - \tilde{\pr}_{0,0} - \tilde{\pr}_{1,1} - \tilde{\pr}_{1,0} - \left(\dfrac{\pr_\ell^A}{\pr_d^A}\right) \pr(D_c,\!D_v \vee D_c,\!D'_\ell \vee D'_c,\!D_\ell \vee D_c D'_\ell,\!D_v \vee D_\ell D'_c,\! D_v \ \! ;  1)\nonumber \\
&&\ \ \ - \left(\dfrac{\pr_\ell^B}{\pr_d^B}\right) \pr(D_v,\!D_c \vee D_\ell,\!D'_c \vee D'_\ell,\!D_c \vee D_v,\!D_c D'_\ell \vee D_v,\!D_\ell D'_c \ \! ;  1) \\
&&\ \ \ -\; 3 \left(\dfrac{\pr_\ell^{A}}{\pr_d^{A}}\right)^{\!\!\!2} \pr(D_c D'_c,\! D_v \ \! ; 1)- 3 \left(\dfrac{\pr_\ell^{B}}{\pr_d^{B}}\right)^{\!\!\!2} \pr(D_v,\!D_c D'_c \ \! ; 1) - 3 \left(\dfrac{\pr_\ell^A \;\! \pr_\ell^B}{\pr_d^A \;\! \pr_d^B}\right) \pr(D_c,\!D'_c \vee D'_c,\!D_c \ \! ; 1).\nonumber
\end{eqnarray}

Note that, to compute $\pr_{key}=\tilde{\pr}_{00} + \tilde{\pr}_{11}$ using the indirect method, we have
\begin{eqnarray}
\tilde{\pr}_{0,0}&&=\pr(D\!,\!R)\! - \pr(D_c,\!R \vee D'_c,\!R \vee D_\ell D'_c,\!R \vee D_c D'_\ell,\!R)\! - \pr(D_c D'_c,\! R)\! -\pr(D_v,\!R \vee D_\ell,\!R \vee D'_\ell,\!R \vee D_\ell D'_\ell,\!R \ \! ; 0)\nonumber\\
&&\ \ \ \ - \; \pr(D_v,\!R \vee D_\ell,\!R \vee D'_\ell,\!R \vee D_\ell D'_\ell,\!R \ \! ; v) -  \pr(D_v,\!R \vee D_\ell,\!R \vee D'_\ell,\!R \vee D_\ell D'_\ell,\!R \ \! ; m), 
\end{eqnarray}
\begin{eqnarray}
\tilde{\pr}_{1,1}&&=\pr(R,\!D)\! - \pr(R,\!D_c \vee R,\!D'_c \vee R,\!D_\ell D'_c \vee R,\!D_c D'_\ell)\! - \pr(R,\!D_c D'_c)\! -\pr(R,\!D_v \vee R,\!D_\ell \vee R,\!D'_\ell \vee R,\!D_\ell D'_\ell \ \! ; 0)\nonumber\\
&&\ \ \ \ - \; \pr(R,\!D_v \vee R,\!D_\ell \vee R,\!D'_\ell \vee R,\!D_\ell D'_\ell \ \! ; v) -  \pr(R,\!D_v \vee R,\!D_\ell \vee R,\!D'_\ell \vee R,\!D_\ell D'_\ell \ \! ; m) .
\end{eqnarray}

From our experimental data, we obtain $\pr_{0,0} = (8 \pm 2) \times 10^{-3}  $, $\pr_{1,1} = (6 \pm 2) \times 10^{-3}$, $\pr_{0,1} = (3 \pm 2) \times 10^{-3} $ and $\pr_{1,0} = (0.5 \pm 2) \times 10^{-3} $. All these values are compatible with those obtained with the direct estimation within experimental uncertainties, which can be reduced by employing a larger sample and improving the single-photon sources and detectors.

\section{Dependence on detection efficiency}\label{sec:detector_eff}

In this section, we discuss the dependence of the secret key rate on the detection efficiencies of Alice's and Bob's detectors, $\pr^A_d$ and $\pr^B_d$, respectively. Note that, since we only consider the first term in Equation~\eqref{eq:cond_entropy} to estimate a bound on $S(A|C)$, we only need to compute the probabilities $\pr_{0,0},\pr_{1,1},\pr_{0,1}, \pr_{1,0}$ and $\pr(1)$. However, $\pr_{0,0},\pr_{1,1},\pr_{0,1},\pr_{1,0}$ are independent of $\pr^A_d$ and $\pr^B_d$, and it is only $\pr(1)$ that has this dependence. Therefore, using the experimental data corresponding to $\pr^A_\ell=1-\pr^A_d=0.42$ and $\pr^B_\ell=1-\pr^B_d=0.42$, we can rewrite $\pr(1)$ with the explicit dependence on the general parameters, $\tilde{\pr}^A_\ell$ and $\tilde{\pr}^B_\ell$, as
\begin{equation}
\begin{array}{ll}
\mathcal N (\tilde{\pr}^A_\ell,\tilde{\pr}^B_\ell)&\!\!= \bk{k_{0,0}} \!+\! \left(\dfrac{\tilde{\pr}^A_\ell}{\pr^A_\ell}\right) \left(\bk{k^1_{0,0}} \!+\! \bk{k^2_{0,0}}\right) \!+\! \left(\dfrac{\tilde{\pr}^A_\ell}{\pr^A_\ell}\right)^{\!\!\!2} \bk{k^3_{0,0}} \\ [4mm]
&\ \ \ + \bk{k_{1,1}} \!+\! \left(\dfrac{\tilde{\pr}^B_\ell}{\pr^B_\ell}\right) \left(\bk{k^1_{1,1}} \!+\! \bk{k^2_{1,1}} \right)\!+\! \left(\dfrac{\tilde{\pr}^B_\ell}{\pr^B_\ell}\right)^{\!\!\!2}\bk{k^3_{1,1}} \\ [4mm]
&\ \ \ + \bk{k_{0,1}}+ \bk{k_{1,0}} + \left(\dfrac{\tilde{\pr}^A_\ell}{\pr^A_\ell}\right)\bk{k^{1}_{0,1}} \!+\! \left(\dfrac{\tilde{\pr}^B_\ell}{\pr^B_\ell}\right)\bk{k^{2}_{0,1}} \\ [4mm]
&\ \ \ + \;\! \left(\dfrac{\tilde{\pr}^A_\ell \;\! \tilde{\pr}^B_\ell}{\pr^A_\ell \;\! \pr^B_\ell}\right)\left(\bk{k^{3}_{0,1}} \!+\! \bk{k^{4}_{0,1}}\right) \!+\! \left(\dfrac{\tilde{\pr}^A_\ell}{\pr^A_\ell}\right)^{\!\!\!2} \bk{k^{5}_{0,1}} \!+\! \left(\dfrac{\tilde{\pr}^B_\ell}{\pr^B_\ell}\right)^{\!\!\!2} \bk{k^{6}_{0,1}} \\ [6mm]
& = \pr (1) (\tilde{\pr}^A_\ell,\tilde{\pr}^B_\ell). 
\end{array}
\end{equation}
Moreover, $\pr_{err}=\pr(1)-\pr_{key}$ is also modified accordingly, to be used in computing $Q = \pr_{err}/\pr(1)$ to obtain the secret key rate presented in Figure~3 from the main paper.

\section{Dependence on transmission loss}\label{sec:losses}

In this section, we provide a brief analysis of the dependence of the key rate on the channel losses, for the case of imperfect photon sources and detectors. The channel loss, after photons passing a distance $L$ through a medium described by the absorption coefficient $\alpha$ (in dB/unit distance), is given by $\ell = \alpha L$. In our protocol, the photons are traveling from the server to the agents, and back, meaning that the total distance $L$ is twice the distance between the server and the agents. This is also the maximal distance between Alice and Bob, achieved when the two are at the opposite sides of the server.

First, note that in general, the all-powerful adversary is bounded only by the laws of physics. In particular, it can vary the number of photons in front of Alice's and Bob's labs at will. But such assumption would seem to turn senseless the whole loss analysis. Moreover, the agents can check the photon number statistics in their labs, thus the adversary must keep them at the levels of the honest case. Finally, note that the overall photon-adversary state in front of the agents has the same shape  as in the lossless case. Indeed, expression \ref{eq:initial_state} represents the most general photon-adversary state that contains up to two photons, in which the probabilities of having zero, one, or two photons are incorporated in the norms of vectors $\ket{c_{a,b}}_C \in \mathcal{H}_C$. 

In our table-top experimental implementation, due to the low transmission loss in air for the considered distance, we can assume that the loss is for all practical purposes zero. Let us fix the source parameters $\pr_1$ and $\pr_2$ (the probabilities of single- and double-photon emission per pulse, respectively), the detector efficiency $\pr_d$ (for simplicity, we assume that the agent's detectors have the same efficiency), and take a certain number of rounds $N$. For that, we can calculate the key rate $r(\ell=0;N)$, presented in Figure 4 from the main text. We have that $N = N_0 + N_1 + N_2$, where $N_i$ is the number of rounds with $i=0,1,2$ emitted photons.

Given the transmission probability $T(\ell) = e^{-\frac{\ell}{10}}$, one can calculate
\begin{eqnarray}\label{eq:rounds_after_losses}
N_0(\ell) & = & N_0 + (1-T)N_1 + (1-T)^2 N_2 \nonumber \\
N_1(\ell) & = & TN_1 + 2T(1-T)N_2 \\
N_2(\ell) & = & T^2 N_2, \nonumber
\end{eqnarray}
where $N_i(\ell)$ are the expected numbers of rounds with $i=0,1,2$ photons present. Note that $N_0(\ell) + N_1(\ell) + N_2(\ell) = N_0 + N_1 + N_2 = N$.

Consider the number of rounds $N' < N$, for which $N_1(\ell) = \pr_1 N'$ and calculate the secret key $r(0;N')$. Then, we have that $r(\ell;N) \geq r(0;N')$, the secret key for $N'$ rounds in the configuration with $L = 0$ is the lower bound of the secret key for $N$ rounds with the loss $\ell$. This bound is based on the following two arguments:
\begin{itemize}
\item[1.] The vacuum pulses neither contribute to the key generation, nor to eavesdropping (they leak no information to the adversary). Thus, only the numbers of single-photon emissions and double-photon emissions are relevant, i.e., whenever we have the key rate for the number of rounds that involve certain numbers of the single-photon and double-photon emissions, we can take this result as valid for any case of having the same single- and double-photon rounds (provided there are no higher-photon rounds).

\item[2.] Given a certain number of rounds, $N$, as $\ell$ grows, both $N_1(\ell)$ and $N_2(\ell)$ decrease. But their ratio does not stay the same, i.e., there exists no $N'$, such that both requirements $N_1(\ell) = \pr_1 N'$ and  $N_2(\ell) = \pr_2 N'$ are satisfied. In other words, the profile of the source changes with $\ell$. But, the ratio $N_1(\ell)/N_2(\ell)$ increases: as $\ell$ grows, there are proportionally more single-photon rounds than double-photon ones, meaning it is more likely that Alice and Bob receive a single photon than two photons. Since double-photon rounds are the ones that, on one side might induce errors in the key, and on the other help the adversary, we actually have that our $r(\ell=0;N')$ is in fact the lower bound for $r(\ell>0;N)$.
\end{itemize}
Thus, having our results $r(N)$ for $\ell=0$, our key rate as a function of the loss $\ell$ is given by
\begin{equation}
	\tilde{r}(\ell) \equiv r(N') = r\Bigg(\frac{N_1(\ell)}{\pr_1}\Bigg),
\end{equation}
where by $\tilde r$ we denote the functional dependence of the key rate on the losses, which is different from the dependence of $r$ on the number of rounds for $\ell=0$. Using the second line of~\eqref{eq:rounds_after_losses}, $N_1 = \pr_1 N$ and $N_2 = \pr_2 N$, we finally have
\begin{equation} 	
\tilde{r} (\ell) = r \Big(\big[ 10^{-\ell/10}\pr_1 + 2\cdot 10^{-\ell/10}\big(1-10^{-\ell/10}\big)\pr_2 \big]N\Big).
\end{equation}

\bibliographystyle{unsrtnat}
\bibliography{manuscript_2}

\end{document}